\documentclass[submission, Phys]{SciPost}

\hypersetup{
    colorlinks,
    linkcolor={red!50!black},
    citecolor={blue!50!black},
    urlcolor={blue!80!black}
}

\usepackage[bitstream-charter]{mathdesign}
\usepackage{braket}
\DeclareSymbolFont{usualmathcal}{OMS}{cmsy}{m}{n}
\DeclareSymbolFontAlphabet{\mathcal}{usualmathcal}

\begin{document}

\begin{center}{\Large \textbf{
Localization control born of intertwined quasiperiodicity and non-Hermiticity
}}\end{center}

\begin{center}
Junmo Jeon, 
SungBin Lee$^\star$
\end{center}

\begin{center}
Korea Advanced Institute of Science and  Technology, Daejeon 34141, South Korea
\\
${}^\star$ {\small \sf sungbin@kaist.ac.kr}
\end{center}

\begin{center}
\today
\end{center}


\section*{Abstract}
{\bf
Quasiperiodic systems are neither randomly disordered nor translationally invariant in the absence of periodic length scales. Based on their incommensurate order, novel physical properties such as critical states and self-similar wavefunctions have been actively discussed.
However, in open systems generally described by the non-Hermitian Hamiltonians, it is hardly known how such quasiperiodic order would lead to new phenomena. 
In this work, we show for the first time that the intertwined quasiperiodicity and non-Hermiticity can give rise to striking effects: perfect delocalization of the critical and localized states to the extended states. In particular, we explore the wave function localization character in the Aubry-Andr\'{e}-Fibonacci (AAF) model where non-reciprocal hopping phases are present. Here, the AAF model continuously interpolates the two different limit between metal to insulator transition and critical states, and the non-Hermiticity is encoded in the hopping phase factors. Surprisingly, their interplay results in the perfect delocalization of the states, which is never allowed in quasiperiodic systems with Hermiticity. By quantifying the localization via inverse participation ratio and the fractal dimension, we discuss that the non-Hermitian hopping phase leads to delicate control of localization characteristics of the wave function. Our work offers (1) emergent delocalization transition in quasiperiodic systems via non-Hermitian hopping phase, 
(2) detailed localization control of the critical states. In addition, we suggest an experimental realization of controllable localized, critical and delocalized states, using photonic crystals.   
 }

\vspace{10pt}
\noindent\rule{\textwidth}{1pt}
\tableofcontents\thispagestyle{fancy}
\noindent\rule{\textwidth}{1pt}
\vspace{10pt}

\section{Introduction}
Quasiperiodic order, which is a novel spatial pattern without any periodic unit length scales, has attracted interest from a wide range of physics disciplines\cite{jaric2012introduction,belin2000quasicrystals,suck2013quasicrystals,janssen2018aperiodic,macia2020quasicrystals}. In a quasiperiodic system, not only the diffraction pattern\cite{jaric1986diffraction,suck2013quasicrystals}, but also the electromagnetic and topological properties would be different from conventional periodic crystals\cite{poon1992electronic,goldman1991quasicrystal,sarkar2021anderson,rosa2021exploring,xiao2021observation,PhysRevResearch.3.013168}. This is mainly due to the localized nature of their quantum states, which decay along a power-law scale and are neither localized nor extended, so-called critical states\cite{deguchi2012quantum,PhysRevB.35.1020,Tuz:09,PhysRevB.96.045138,PhysRevLett.58.2436,PhysRevB.101.174203,PhysRevB.107.054206,aubry1980analyticity,Shen1994}. The critical states arise from the incommensurate self-similar quasiperiodic ordered structure\cite{PhysRevB.96.045138,PhysRevLett.58.2436}. Theoretically, the quasiperiodic systems and possible critical states in one-dimensional chains have been actively studied in terms of the Aubry-André model\cite{aubry1980analyticity,PhysRevLett.115.180401} and the Fibonacci quasicrystal\cite{PhysRevB.96.045138,PhysRevB.35.1020}. The critical states that emerge in these systems lead to the stable fractal magnon transmittance which has advanced the field of magnonics\cite{PhysRevLett.97.026601,PhysRevX.6.011016,PhysRevB.106.134431}.
 In recent years, it has become possible to artificially create quasicrystalline structures in the laboratory, such as metamaterials and photonic crystals\cite{PhysRevB.102.075107,PhysRevLett.126.145501,PhysRevB.88.201404,PhysRevResearch.4.043030,PhysRevLett.125.200604,stadnik1998physical} and the potential for experimental applications using quasiperiodicity and critical states has been increased. However, how the quasiperiodic orders and their critical states behave in the open systems is still poorly understood.

To understand open systems where energy is not conserved, the effective non-Hermitian theory has been used as a theoretical framework by neglecting quantum jumps from the standard Lindblad equation\cite{PhysRevX.13.021007,10.1063/1.5115323,niu2023effect}. In the last decade, there has been a growing interest in the systems described by non-Hermitian Hamiltonians. For example, non-Hermitian Hamiltonians have been widely used in exciton-polariton theory\cite{gao2015observation,PhysRevLett.125.123902,PhysRevB.104.235408}, photonics\cite{wang2021topological}, and other optical systems\cite{carmichael2009open}. More recently, researchers have developed non-Hermitian descriptions of magnonic systems\cite{PhysRevB.105.L180406,hurst2022non}. These have been used to describe magnons in driven and dissipative spintronic systems\cite{hurst2022non}. There is also a non-Hermitian tight-binding model with the non-reciprocal hopping magnitudes that exhibits the non-Hermitian skin effect, the massive condensation of bulk modes to the edge under the open boundary condition due to the nonzero spectral area under the periodic boundary condition\cite{PhysRevX.13.021007,zhang2022universal,PhysRevB.104.125416,PhysRevB.100.054301,PhysRevB.103.054203,PhysRevResearch.3.033184,PhysRevLett.77.570,PhysRevLett.80.5172,PhysRevLett.124.086801,zhang2021observation,zhou2022engineering,PhysRevLett.127.256402,yao2018edge,PhysRevB.106.134112,doi:10.1126/science.abf6568}. 
Such non-Hermitian systems have been realized experimentally using photonic crystals, fiber optics, or electrical circuits\cite{doi:10.1126/science.aaz3071,wang2021topological,lee2018topolectrical,Wu:16,PhysRevA.102.023501,gu2022transient}. 
Furthermore, the effect of non-Hermiticity in quasiperiodic systems has recently been spotlighted in terms of the non-Hermitian skin effect, parity-time symmetry and the topological phase transitions \cite{PhysRevB.100.054301,PhysRevB.100.125157,PhysRevB.104.224204,PhysRevB.106.014204,PhysRevResearch.2.033052,PhysRevLett.122.237601,PhysRevB.103.104203}. Nonetheless, it remains to be understood how such non-Hermiticity could be used to manipulate drastic changes of the quantum states in quasiperidic systems.

\begin{figure*}[t]
\centering
\includegraphics[width=0.9\textwidth]{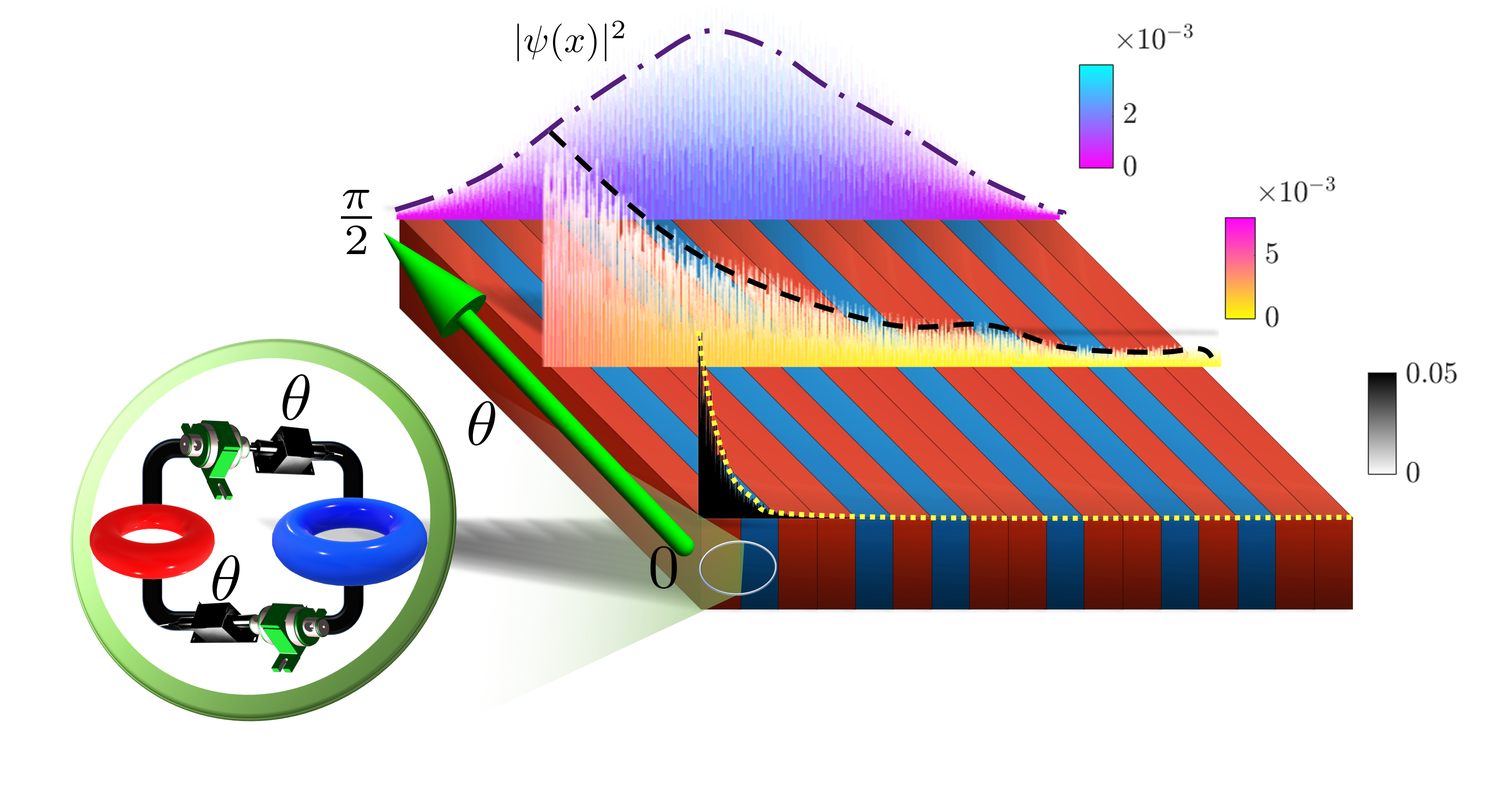}
\caption{\label{fig: impact} A proposal for an experimental control of localization characteristics in non-Hermitian optical system with ring resonators. A simplified representation of the ring resonators that make up the quasiperiodic pattern is shown in the sequence of red and blue blocks. Two or more ring resonators with different resonance frequencies are arranged in a quasi-periodic fashion to form a photonic crystal. Here, the ring resonators which have two different resonance frequencies are drawn as red and blue rings in the zoomed inset. Adjacent ring resonators are connected to each other by two optical cables with optical isolator and phase shifter, depicted as the green and black components in the zoomed inset, respectively. Each optical isolator allows the transmission of light in left- and right-moving light only, respectively, and the phase shifter adds the same phase on the transmitted wave. Thus, we have effective non-reciprocal hopping phase for the light. By arranging the ring resonators as the Fibonacci quasicrystal, and manipulating the non-reciprocal hopping phase, $\theta$ accumulated by the phase shifters, we can change the localization characteristics of the wave function, $\psi(x)$ from exponentially localized (black) to extended (blue) exploring critical states (yellow) in terms of $\theta$. Two different decaying behaviors of the localized and critical states are drawn by the dotted and dashed lines indicating the envelopes of the probability distributions, respectively. One could measure the optical conductivity to read off the change of the light localization characteristics.}
\end{figure*}

In the current work, we study the system where both quasiperiodic order and non-Hermiticity play an important role and discuss the striking result due to their interplay. In contrast to the traditional non-Hermitian models explained by the Hatano-Nelson argument\cite{PhysRevLett.77.570}, here the non-Hermiticity is taken into account as the non-reciprocal phase of the hopping parameters, instead of the hopping magnitudes.  
As a representative example of the quasiperiodic order, we study the Aubry-André-Fibonacci model\cite{PhysRevResearch.3.033257},  but note that our argument is generally applicable to other types of quasiperiodic systems.
 We show that the interference arising from the non-reciprocal hopping phases and the exceptional coalescence of the states controls the localization of the wave function. Surprisingly, intertwined quasiperiodicity and non-Hermiticity gives rise to a perfect delocalization without fractality of the wave function, which never occurs in a quasiperiodic system with Hermiticity. Not only the emergence of delocalization, but also their interplay leads to the control of critical states, implying potential applications. 

Fig.\ref{fig: impact} illustrates our main results with sketch of potential experimental implications. 
Here, as an example of the quasiperiodic system, the two distinct ring resonators (red and blue blocks) are arrayed in Fibonacci quasicrystalline patterns. Each ring resonator is connected by the optical isolators\cite{trefil2001encyclopedia} and phase shifters which in principle controls the non-reciprocal hopping phase, $\theta$. (See the inset of Fig.\ref{fig: impact}.) By changing the non-Hermiticity given by non-reciprocal hopping phase, $\theta$ from 0 to $\pi/2$, we suggest that one can explore general localization characteristics of the wave functions from the localized (black) to critical (yellow) and extended ones (blue), as represented on the top of the quasiperiodic arrays. Therefore, the non-Hermitian quasiperiodic system can be used to manipulate quantum transport experiments. 

The remainder of the article is organized as follows. In Sec.\ref{sec:model} we describe our non-Hermitian tight-binding Hamiltonian with non-reciprocal hopping phase. In Sec.\ref{sec:delocalization}, by exemplifying the Aubry-André-Fibonacci model, we analyze the change of the localization strength of the state, which is given by the inverse participation ratio in terms of the non-reciprocal hopping phase. We show that the interference effect originated from the intertwined quasiperiodic potential and non-reciprocal hopping phase gives rise to the change of the localization strength of the states. Also, we show that the states are delocalized in the Fibonacci chain as the function of non-reciprocal hopping phase. We explain that the exceptional hybridization of the eigenstates leads to the change of the localization characteristics of the states. In Sec.\ref{sec:discussion}, we summarize our works.
\section{Non-Hermitian Hamiltonian with non-reciprocal hopping phase}
\label{sec:model}
Let us consider the tight-binding Hamiltonian with $N$-sites
\begin{align}
\label{hamiltonian}
&H=H_V+H_T, \\
&H_V=\sum_{i=1}^NV_i\ket{i}\bra{i}, \nonumber \\
&H_T=t\sum_{i=1}^{N-1}(\ket{i}\bra{i+1}+\ket{i+1}\bra{i}), \nonumber \\
&t=Te^{i\theta}, \nonumber
\end{align}
where $V_i$ are non-uniform real-valued local potentials. $t$ is the uniform complex-valued hopping parameters, respectively. $\ket{i}$ represents the particle placed on the $i$-th site. 
The hopping parameters are uniform complex-valued, $t=Te^{i\theta}$ where $T$ and $\theta$ are positive reals. The strength of the non-Hermiticity is given by $T\sin{\theta}$, which is maximized when $\theta=\pi/2$. Note that the strength of the non-Hermiticity is depending on both hopping magnitude and the non-reciprocal hopping phase. We will show that the non-Hermiticity encoded in the non-reciprocal hopping phase changes the localization properties of the states compared to the Hermitian counterparts given by $\theta=0,\pi$.

One of the most important quantity used in the non-Hermitian systems is the phase rigidity, which is defined by
\begin{align}
\label{phaserigidity}
&r(\psi_k)=\vert\braket{\psi_k^{(L)}|\psi_k^{(R)}}\vert.
\end{align}
Here, the superscripts $L$ and $R$ stand for left and right eigenstates of the non-Hermitian Hamiltonian, and the subscript $k$ is the index of eigenstate. Unlike the Hermitian systems where the phase rigidity is always 1, it could be less than one, and even vanished in the non-Hermitian systems. Particularly, when two distinct eigenstates coalesce, the phase rigidity becomes zero\cite{PhysRevX.6.021007,PhysRevA.95.022117}. This unique characteristics of the non-Hermitian system is called an exceptional point. Thus, one can use the phase rigidity to quantify the coalesence of the states in the non-Hermitian system.

We quantify the localization strength of the state by using the inverse participation ratio (IPR), which is defined for a normalized state $\psi$ as
\begin{align}
\label{ipr loc}
\mbox{IPR}(\psi)=\sum_i|\psi(i)|^4.
\end{align}
Note that the amount of localization for the wave function, $\psi$, can be quantified by the IPR\cite{PhysRevB.83.184206,calixto2015inverse,PhysRevB.100.054301}. In the spectrum, the maximum (minimum) value of the IPR indicates the maximally (minimally) localized state. Let us refer to these states in the spectrum as maximally localized and maximally extended states, respectively. Also, the average localization strength for entire states in the spectrum is given by the mean IPR (MIPR), defined by
\begin{align}
\label{MIPR}
\mbox{MIPR}=\frac{1}{N}\sum_{k=1}^{N}\mbox{IPR}(\psi_k),
\end{align}
where $\psi_k$ is the $k$-th eigenstate. The delocalization (localization) can be captured by the reduction (enhancement) of MIPR\cite{PhysRevB.100.054301}.



\section{Delocalization in the non-Hermitian quasiperiodic chains}
\label{sec:delocalization}
Given a finite hopping magnitude, $T$, we can change the localization characteristics of the states as $\theta$ approaches $\pi/2$, where the strength of non-Hermiticity becomes maximum. To understand how the non-Hermitian hopping phase factor allows to change the localization of the states, let us consider the return probability, which measures the probability that the particle placed at the $i$-th site will return to the $i$-th site. A smaller return probability indicates that the state is more delocalized because the wave function is scattered. Using the path integral idea, the return probability is given by the sum of the transition amplitudes from all possible paths whose start and end points are the same as the $i$-th site. If the hopping phase is non-reciprocal, there is destructive interference between the transition amplitudes. This interference originated from the non-reciprocal hopping phase essentially gives rise to the delocalization of the state by reducing the return probability. On the other hand, the non-reciprocal phase can also lead to the constructive interference with respect to $\theta$ for some states. In this case, the delocalization is hindered by the interference from the non-reciprocal hopping phase. Thus, the non-reciprocal hopping phase gives rise to the \textit{state-dependent} control of the localization properties. Moreover, in the non-Hermitian systems, such interference effect leads to the unconventional coalesence of the states which have different localization characteristics, so-called exceptional points. Hence, before and after this exceptional point, the localization characteristics would be changed drastically as we will demonstrate.

To illustrate the \textit{state-dependent} control of the localization strength with a concrete argument, we consider a toy model consisting of two different atoms, $A$ and $B$, arranged in an alternating way, i.e. $ABABAB\cdots$. This can be understood as the $1/1$-approximant of the Fibonacci chain, which we discuss in Sec.\ref{sec:Fibonacci}\cite{Shen1994}. With periodic boundary conditions, the Hamiltonian in momentum space is given by the $2\times 2$ matrix below
\begin{align}
\label{toymodel}
&H(k)=\begin{pmatrix} V_A & t(1+e^{-ika}) \\ t(1+e^{ika}) & V_B \end{pmatrix}
\end{align}
where $a$ is the distance between the atoms, $k$ is the momentum, and $V_A$ and $V_B$ are the local potentials of the $A$ and $B$ atoms, respectively. The eigenvectors given by
\begin{align}
\label{toymodel2}
&\ket{v_\pm(k)}=\frac{1}{\mathcal{N}_\pm} \begin{pmatrix} 2t(1+e^{-ika}) \\ \Delta V\pm\sqrt{\Delta V^2+16t^2\cos^2{ka/2}} \end{pmatrix}
\end{align}
where $\Delta V=V_B-V_A$ which is assumed to be positive without loss of generality. $\mathcal{N}_\pm$ is a normalization constant. In the eigenstates, $\Delta V^2+16t^2\cos^2{ka/2}$ indicates the interference between the potential difference and hopping contributions. Note that this interference effect is the result of the local potential gradient, $\Delta V$ and the non-reciprocal hopping phase $\theta$ in $t=Te^{i\theta}$. Furthermore, the relative probability amplitude at $B$ sublattices is reduced (increased) as the function of $\theta$ for $v_{+(-)}(k)$. This is because the return probability for the $B$ atoms of $v_{+(-)}(k)$ is reduced (enhanced) due to the state-dependent phase difference between $\Delta V$ and $\sqrt{\Delta V^2+16t^2\cos^2{ka/2}}$. This is a typical example of the state-dependent control of localization characteristics as the result of the interplay between the unflatten potential distribution and the non-Hermiticity.

The result of the interference effect between the non-uniform potential and the non-reciprocal hopping phases, can be understood with the effective potential distribution, $V_i^{\mbox{eff}}$.
 Here, $V_i^{\mbox{eff}}$ is the deformed on-site potential which leads to the same probability distribution of the given state assuming $\theta=0$. In general, the effective potential is the state-dependent, distinguishing between different momentum $k$ from Eq.\eqref{toymodel2}.
Surprisingly, however, when the hopping magnitude $T$ is larger than $\frac{\Delta V}{4\cos(Ka/2)}$ for a given momentum $K$, the probability amplitude at the $B$ sublattices of $v_\pm(k\le K)$ and their localization strengths are saturated at $\theta=\pi/2$ as the same value regardless of the sign difference in Eq.\eqref{toymodel2} and momentum $k\le K$. Thus, the effective potential becomes uniformly periodic for every $v_{\pm}(k\le K)$. In particular, in this case the probability distribution for $k\le K$ becomes uniform, and thus indistinguishable from a uniformly periodic chain with no potential gradient. This shows that if the finite hopping magnitude is sufficiently large compared to the potential difference, the effect of the potential gradient would be washed out by the interference effect arising from the non-reciprocal hopping phase. In other words, the non-reciprocal hopping phase deforms $V_i$ to the uniform effective potential $V^{\mbox{eff}}_i$ such that $\Delta V^{\mbox{eff}}=0$. Such a delocalization of the states would be achieved delicately for each momentum $k$ by increasing $T$. Thus, a non-reciprocal hopping phase provides a high controllability of the localization strength for each state.

Note that when $T=\frac{\Delta V}{4\cos(Ka/2)}$, $v_+(K)$ coalesces into $v_-(K)$, and hence the phase rigidity defined in Eq.\eqref{phaserigidity} for $v_\pm(K)$ becomes zero. Thus, the unconventional coalesence of the states due to the non-Hermiticity is happened when the state is uniformly delocalized. It turns out that the vanishment of the phase rigidity indicates the delocalization transition.

\begin{figure}[]
\centering
\includegraphics[width=0.7\textwidth]{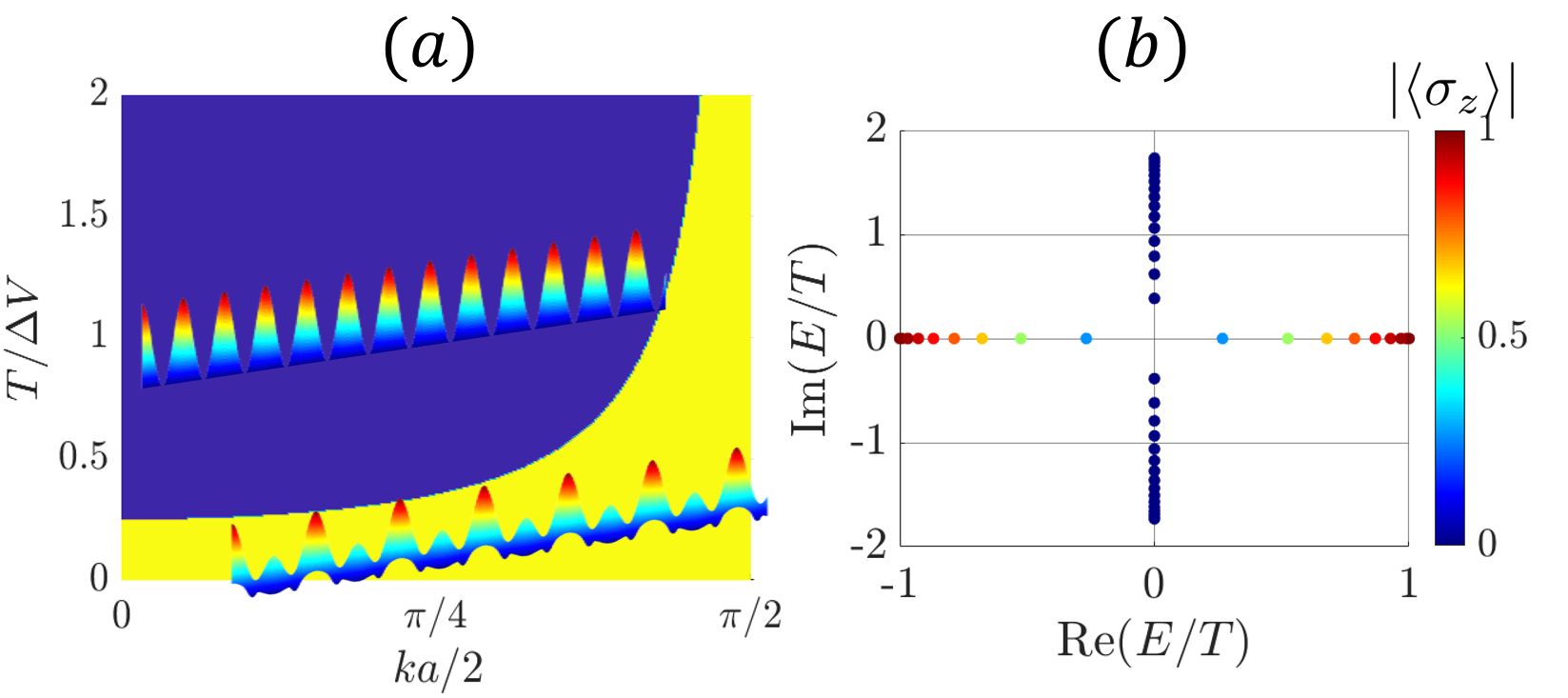}
\caption{\label{fig: toymodel} (a) State-dependent localization transition at $\theta=\pi/2$ for the alternating periodic chain. For each value of $k$, the blue region indicates that the wave function is uniformly distributed regardless of the atomic species, the same as for the uniformly periodic chain. Yellow region indicates that the wave function has skewed probability amplitude between $A$ and $B$ atoms. The probability distributions for each case are drawn schematically. For the Hermitian case, the blue region does not appear for finite $T$. (b) $\left\vert\braket{\sigma_z}\right\vert$ of the eigenstates on the complex energy plane at $\theta=\pi/2$.  Here, we set $T=\Delta V$ and $V_A=-V_B$. The zero $\left\vert\braket{\sigma_z}\right\vert$, which indicates the uniformly distributed wave function, appears with the pure imaginary energies.}
\end{figure}

Fig.\ref{fig: toymodel} (a) shows the phase diagram depending on the presence and absence of the $\Delta V$ effect on the state at $\theta=\pi/2$. Here the order parameter is given by $\left\vert\braket{\sigma_z}\right\vert$, where $\sigma_z$ is the Pauli matrix acting on the wave function of the $A$ and $B$ sublattices in Eq.\eqref{toymodel2}. Note that the order parameter $\left\vert\braket{\sigma_z}\right\vert$ measures the difference between the probabilities on the $A$ and $B$ sublattices. The yellow region indicates that the probability distribution for $A$ and $B$ atomic species is different due to the potential difference, $\Delta V$. In contrast, the blue region indicates that for each $k$ the probability distribution of $v_\pm(k)$ is completely uniform and independent of the atomic species. The phase boundary is given by the zero phase rigidity for each $k$ values. Note that when $\theta=0$, hermitian system, the blue region is impossible for any $k$ and finite $T$. This is because the non-reciprocal hopping phase could induce the delocalization of the states by interference effect instead of just increasing the mobility of the particle. Fig.\ref{fig: toymodel} (b) illustrates the order parameter, $\left\vert\braket{\sigma_z}\right\vert$ on the complex energy plane at $T=\Delta V$ and $\theta=\pi/2$. Remarkably, the zero values of $\left\vert\braket{\sigma_z}\right\vert$ appear only for the complex valued energies. This is because the energy eigenvalues of the Hamiltonian, Eq.\eqref{toymodel} are given by 
\begin{align}
\label{toymodelenergy}
&E_{k,\pm}=\frac{1}{2}(V_A+V_B\pm\sqrt{\Delta V^2+16t^2\cos^2(ka/2)}).
\end{align}
Hence, at $\theta=\pi/2$, the energy becomes complex if and only if $T$ is larger than $\frac{\Delta V}{4\cos(ka/2)}$, which is the blue region of Fig.\ref{fig: toymodel} (a).

In terms of the hopping phase, one can explore different localization characteristics, not only exponentially localized or uniformly extended, but also power-law decaying critical states depending on $V_i$. To show the high controllability of the localization characteristics of the states, we consider the quasi-periodic systems in one-dimensional space. We study the Aubry-André Fibonacci model and discuss the effect of non-Hermiticity induced by non-reciprocal hopping phase. This model includes the Fibonacci quasicrystal limit, where both exponentially localized and critical wave functions exist. We emphasize that our discussion can be generalized to other systems or to the randomly disordered systems \cite{https://doi.org/10.48550/arxiv.2112.14783,PhysRevB.105.064502,https://doi.org/10.48550/arxiv.2205.15343} (see Appendix \ref{sec: A1} for detailed information).



\subsection{Aubry-André-Fibonacci model}
Aubry-André-Fibonacci (AAF) model\cite{PhysRevResearch.3.033257} is 1D chain with quasi-periodically modulated $V_i$ given by,
\begin{align}
\label{AAF potential}
V_i(\beta)=-\lambda\frac{\tanh[\beta\cos(2\pi\alpha i+\varphi)-\beta\cos(\pi\alpha)]}{\tanh\beta}.
\end{align}
Here $\varphi$ is the phase shift, indicating the global spatial translation of the potential, which we set to $\varphi=0$. $\alpha$ is the golden section, $(1+\sqrt{5})/2$. With respect to $\beta$, the model deforms continuously from the Aubry-André model\cite{aubry1980analyticity} ($\beta\to0$) to the Fibonacci limit\cite{PhysRevResearch.3.033257} ($\beta\to\infty$), which we discuss in Sec.\ref{sec:Fibonacci}. The Aubry-André model has been actively studied for the metal-insulator transition with respect to $T/\lambda$ in the incommensurate ordered systems\cite{PhysRevResearch.3.033257,PhysRevB.100.125157,PhysRevB.106.014204}.
Given $\beta$, we investigate the localization strength of the states as a function of the non-reciprocal phase, $\theta$ and $T/\lambda$.

Before illustrating the results, we briefly review the localization properties of the Hermitian AAF model under the open boundary condition (OBC). Under OBC, the Hermitian AAF model possesses at least one exponentially localized state regardless of finite $T$ and $\lambda\neq 0$. Thus, the maximally localized state in AAF models is exponentially localized regardless of $T/\lambda$ and $\beta$\cite{PhysRevB.104.014202}. For example, when $\beta=0$, the fraction of localized states among the eigenstates changes drastically at the boundary of $T/\lambda=1/2$ from $1$ to $\sim 1/N$, where $N$ is the finite system size. This change is known as the metal-insulator phase transition, where the insulating and metallic phases refer to $T/\lambda<1/2$ and $T/\lambda>1/2$ regimes, respectively\cite{aubry1980analyticity}. For nonzero $\beta$, there are only a few delocalized states for $T/\lambda<1/2$. For example, at $\beta=2.5$, the delocalized state exists even for $T=0.05\lambda$.

\begin{figure}[]
\centering
\includegraphics[width=0.7\textwidth]{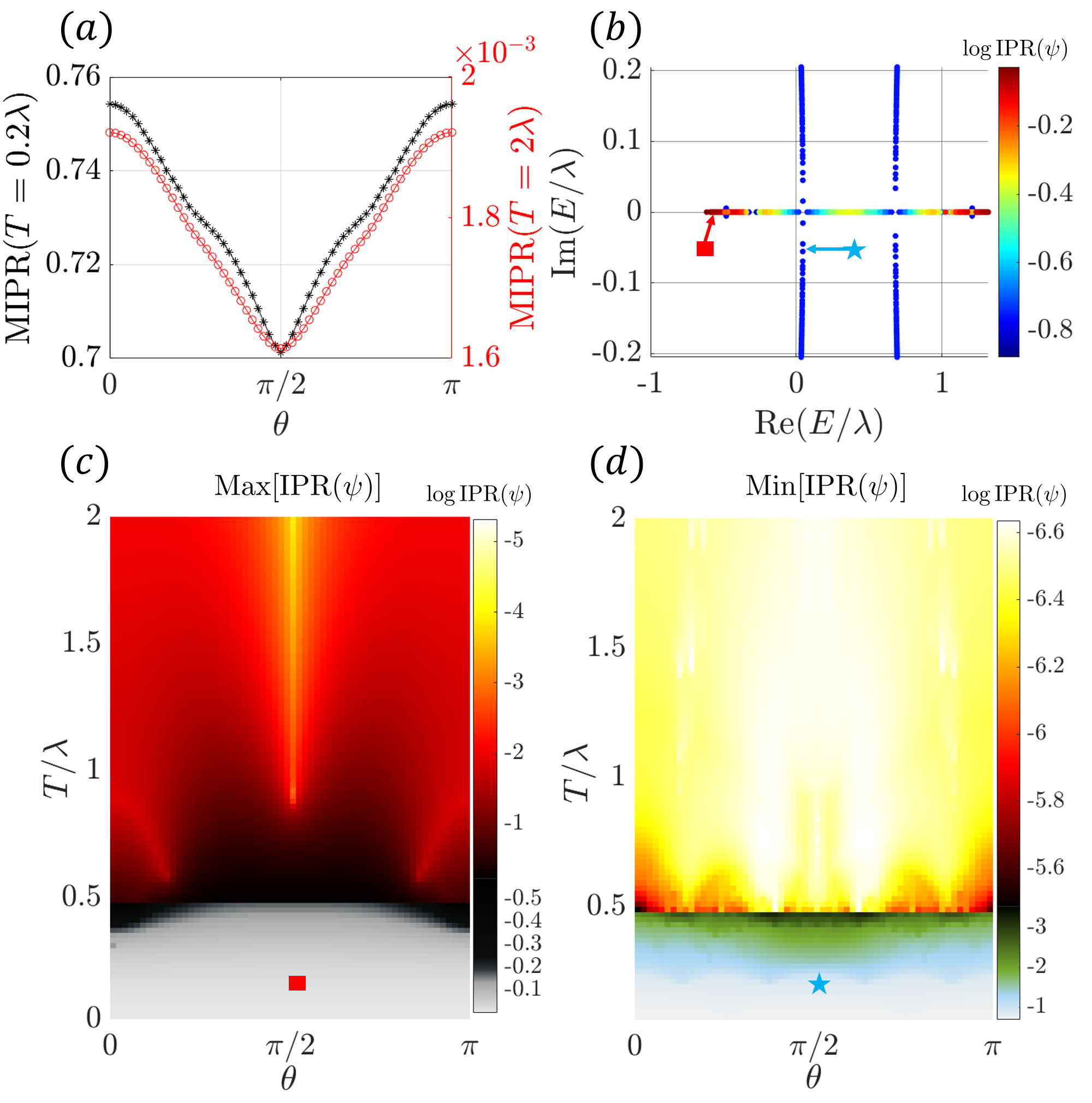}
\caption{\label{fig: zerobeta} Changes of localization strength in the non-Hermitian AAF model for $\beta=0$ as a function of $\theta$. 
(a) Suppression of MIPR as $\theta$ approaches $\pi/2$, for both $T=0.2 \lambda$ (black curve) and $T=2\lambda$ (red curve). 
(b) $\log(\mbox{IPR})$ of the states on the complex energy plane at $T=0.2\lambda$ and $\theta=\pi/2$. Smaller $\log(\mbox{IPR})$ values appear mostly with the complex-valued energies. The red square and blue star indicate the maximally localized and maximally extended states in (c) and (d), respectively. Landscape of $\log(\mbox{IPR})$ for (c) maximally localized state and (d) maximally extended state. For $T<0.5\lambda$, (c) the localization strength of the maximally localized state increases in terms of IPR enhancement as $\theta$ gets closer to $\pi/2$. (d) The maximally extended state shows the decrease of the IPR as a function of $\theta$. (c,d) In contrast, for $T\ge0.5\lambda$, both maximally localized and maximally extended states are delocalized as $\theta$ approaches $\pi/2$.
}
\end{figure}

\begin{figure}[]
\centering
\includegraphics[width=0.7\textwidth]{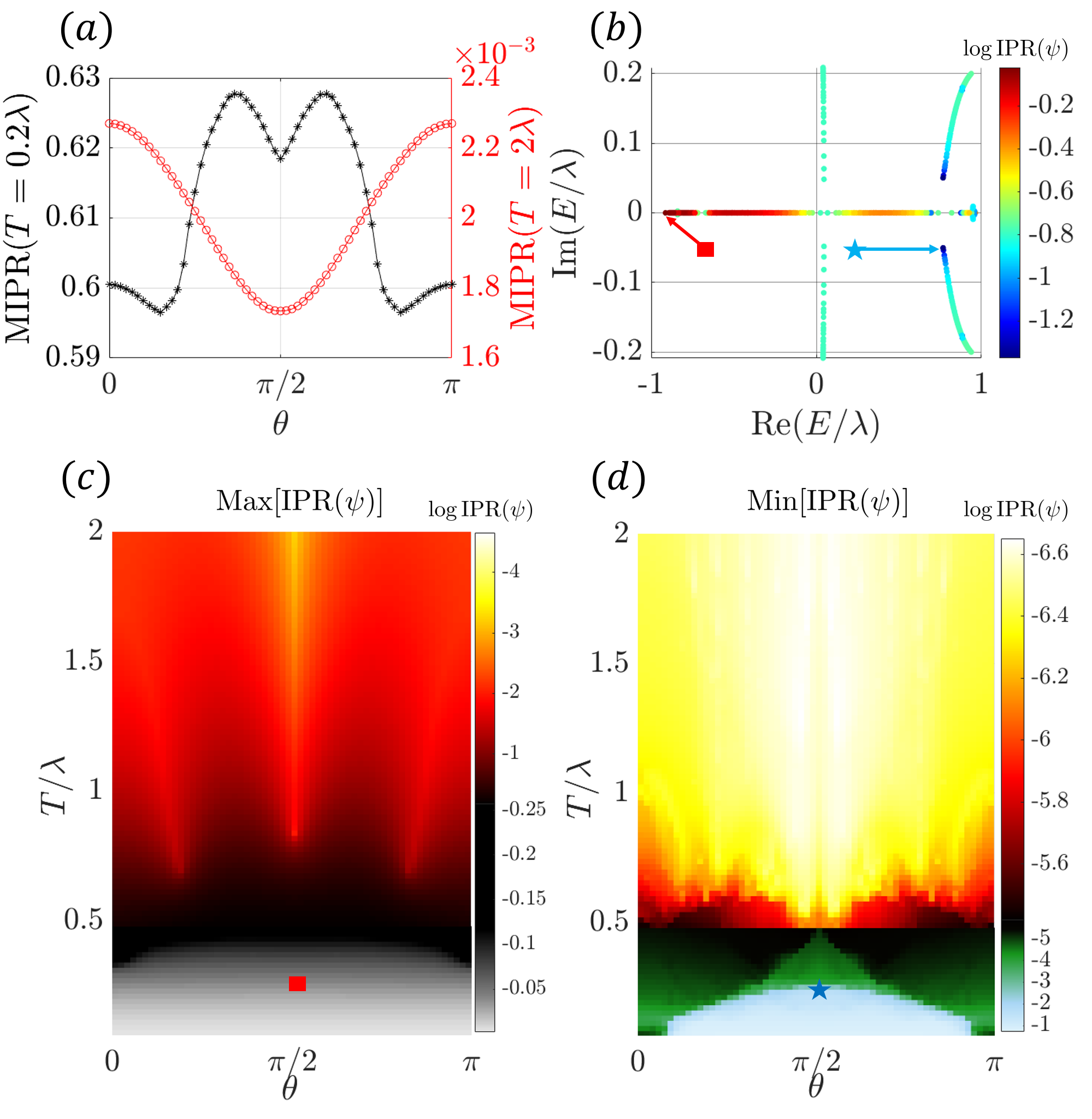}
\caption{\label{fig: nonzerobeta} Changes in localization strength in the AAF model for $\beta=2.5$  as a function of $\theta$. (a) The MIPR for small hopping magnitudes increases for non-trivial $\theta$, as shown in black curve for $T=0.2\lambda$. Whereas, it monotonically decreases for large hopping as $\theta$ approaches $\pi/2$, as presented in red curve for $T=2\lambda$. (b) $\log(\mbox{IPR})$ of the states on the complex energy plane at $T=0.2\lambda$ and $\theta=\pi/2$. The states having large $\log(\mbox{IPR})$ values admit the real-valued energies, while the states with relatively small $\log(\mbox{IPR})$ values have the complex-valued energies. The red square and blue star indicate the maximally localized and maximally extended states in (c) and (d), respectively. Landscape of $\log(\mbox{IPR})$ for (c) maximally localized state and (d) maximally extended state. For $T<0.5\lambda$, (c,d) the localization strengths of the maximally localized and  maximally extended states increase in terms of IPR enhancement as $\theta$ approaches $\pi/2$. In particular, the sky region in (d) indicates the exponentially localized states. (c,d) In contrast, for $T\ge0.5\lambda$, both maximally localized and maximally extended states are delocalized as $\theta$ approaches $\pi/2$, similar to the case of $\beta=0$.
}
\end{figure}

Now, we explore the non-Hermitian AAF model, where non-Hermiticity is induced by the non-reciprocal hopping phase $\theta$. For different $\beta$, Fig.\ref{fig: zerobeta} and Fig.\ref{fig: nonzerobeta} represent the IPR change  and illustrate the localization and delocalization as a function of $\theta$.
First of all, for $\beta=0$, Fig.\ref{fig: zerobeta} (a) shows that the MIPR decreases as $\theta$ gets closer to $\pi/2$ regardless of $T$, and thus the delocalization of the state is generally observed. Fig.\ref{fig: zerobeta} (b) shows each state on the complex energy plane and their IPR at $T=0.2\lambda$ and $\theta=\pi/2$. Note that when the IPR of the state is relatively large, the energy of the state remains real-valued, while the energies of the states with smaller IPR values are mostly complex-valued. This is because the non-Hermiticity is driven by non-reciprocal hopping phase, which does not affect on the energy of strongly localized states. 
Here, the maximum and minimum values of the IPR are emphasized as the red square and blue star, respectively (See Fig.\ref{fig: zerobeta} (b-d)). 
Figs.\ref{fig: zerobeta} (c) and (d) illustrate that $T<0.5\lambda$ and $T\ge 0.5\lambda$ show different behavior in the localization strength for each state. In particular, for $T<0.5\lambda$, the IPR of the maximally localized state increases as $\theta$ approaches $\pi/2$ as shown in Fig.\ref{fig: zerobeta} (c). Whereas, Fig.\ref{fig: zerobeta} (d) shows that, for $T<0.5\lambda$,  the IPR of the maximally extended state decreases as $\theta$ gets closer to $\pi/2$. Thus, the change in the localization strength is different depending on the states. This observation supports that the change of the localization strength is different for each state when the strength of the non-Hermiticity, $T\sin\theta$ is small. 
On the other hand, for $T\ge0.5\lambda$, the non-Hermiticity generally leads to the delocalization of each state. Fig.\ref{fig: zerobeta}(c) and (d) show that both the maximum and minimum values of the IPR decrease as $\theta$ approaches $\pi/2$.
This indicates that the states are delocalized due to the non-reciprocal hopping phase. In particular, although the maximally localized state is still exponentially localized for $T\ge0.5\lambda$ under OBC, its localization strength given by the IPR decreases as a function of $\theta$, unlike the case of $T<0.5\lambda$.

Next we consider the case of nonzero $\beta$, exemplifying $\beta=2.5$, which shows significant changes in the localization strength in the $T<0.5\lambda$ regime, compared to the case of $\beta=0$. We observe the state localization with respect to $\theta$ when $T<0.5\lambda$. The black curve of Fig.\ref{fig: nonzerobeta}(a) shows that the MIPR for $T=0.2\lambda$ is enhanced with non-trivial $\theta$. This enhancement of the MIPR is quite general for the regime $T<0.5\lambda$.
In detail, Fig.\ref{fig: nonzerobeta}(c) and (d) show the enhancement of the IPR of the maximally localized and extended states as $\theta$ approaches $\pi/2$ for small $T<0.5\lambda$ regime. 
In particular, Fig.\ref{fig: nonzerobeta}(c) shows that the extended states disappear in the spectrum due to the non-reciprocal hopping phase. Thus, each state is exponentially localized, similar to the case of $\beta=0$. It is because when $T$ is small, the modulation of the potential distribution due to the hopping phase is weak, and hence the effective potential distribution resulting from the interference effect is similar to the case of $\beta=0$ rather than uniformly periodic. Fig.\ref{fig: nonzerobeta} (b) illustrates the $\log(\mbox{IPR})$ on the complex energy plane for $T=0.2\lambda$ and $\theta=\pi/2$. Here, the red square and blue star indicate the maximally localized and extended states. The strongly localized states have the real-valued energies, while the less localized states have the complex-valued energies. This supports that the localization for the $T<0.5\lambda$ regime is strongly affected by the non-reciprocal hopping phase.
For $T\ge 0.5\lambda$, on the other hand, the red curve in Fig.\ref{fig: nonzerobeta} (a) shows the delocalization of the states in terms of MIPR suppression as $\theta$ approaches $\pi/2$. Note that for $T\ge0.5\lambda$ regime, the delocalization of the states is a common signature with the $\beta=0$ case (see Figs.\ref{fig: nonzerobeta} (c) and (d)).


Comparing Fig.\ref{fig: zerobeta} and Fig.\ref{fig: nonzerobeta}, one can conclude that the non-reciprocal hopping phase can either increase or decrease the localization strength depending on the state and the potential distribution dependent on $\beta$. This implies that the non-reciprocal hopping phase allows for different control of the localization strength. Nevertheless, we find that when the hopping magnitude is sufficiently strong, the delocalization of the states is generally induced. This is because the non-Hermitian interference effect which leads to the effective potential is uniform, and hence for $T$ greater than some critical hopping magnitude, say $T_c$, the interference in terms of $\theta$ uniformly washes out the effects of the potential gradient on the probability distribution, as we demonstrated in the example of an alternating periodic chain.

The general delocalization of states in AAF models for large $T$ could be also understood in terms of state hybridization between delocalized states. When $T$ is large, the Hermitian Hamiltonian possesses delocalized states whose probability distribution is locally non-uniform. Then, the non-reciprocal hopping phase induces the hybridization between these delocalized states, which compensates the local difference of probability amplitudes. For example, in the case of an alternating periodic chain, the difference of the probability amplitude at the $A$ and $B$ sublattices in $\ket{v_+(k)}$ and $\ket{v_-(k)}$ is compensated as $\theta$ approaches $\pi/2$. In particular, after the coalescence of $\ket{v_+(k)}$ and $\ket{v_-(k)}$ at $\theta=\pi/2$, the local probability difference between $A$ and $B$ sites is fully compensated. The coalescence of states is a unique feature of the non-Hermitian system\cite{PhysRevX.6.021007,doi:10.1126/science.aar7709}. Thus, such anomalous state hybridization is another key mechanism leading to the delocalization of states. Moreover, this kind of delocalization of states is observed from the weak localization in randomly disordered systems. In particular, for sufficiently large finite $T$, as $\theta$ approaches $\pi/2$, the MIPR decreases, and hence the delocalization of the state is induced (see Appendix \ref{sec: A1} for detailed information).

The non-reciprocal hopping phase can control the localization strength of the states of AAF models in different ways, either increasing or decreasing the localization strength. This opens up new experimental applications of AAF models. For example, based on the result of the case of $T<0.5\lambda$ for $\beta=2.5$, one can control the localization characteristics from the mixture of extended and localized states to the perfectly localized spectrum. On the other hand, when $T>0.5\lambda$, the non-Hermiticity induces the delocalization of the states. Thus, depending on the hopping magnitude, one can either promote or suppress the mobility of the particle as a function of the non-reciprocal hopping phase.

\subsection{Fibonacci quasicrystal}
\label{sec:Fibonacci}
Now let us consider the $\beta\to \infty$ limit of the AAF model, the Fibonacci quasicrystal\cite{fujiwara2007quasicrystals,stadnik1998physical}. In detail, the Fibonacci quasicrystal consists of two different atoms, $A$ and $B$, which have onsite potentials $V_A$ and $V_B$, respectively. From Eq.\eqref{AAF potential}, $V_A=\lambda$ and $V_B=-\lambda$. By using the successive substitution maps, $A\to AB$ and $B\to A$, one obtains the Fibonacci arrangement of the atoms, such as $ABAABABABAA\cdots$\cite{PhysRevResearch.3.013168,PhysRevB.106.134431}. For the Hermitian system under OBC, it is known that there are both exponentially localized states and critical states in the Fibonacci quasicrystal, independent of the finite hopping magnitude, $T$\cite{jullien2012universalities,forrest2002topological,PhysRevB.104.224204}. Thus, one could ask whether non-Hermiticity with a uniform complex-valued hopping parameter enhances the delocalization of the states in the Fibonacci quasicrystal and eventually gives rise to the extended states for a finite hopping magnitude under the OBC.

To quantify the localization characteristics of the wave function, we use both the IPR and the fractal dimension of the state. Although a larger IPR indicates stronger localization, the value of the IPR alone is insufficient to determine the detailed localization characteristics and scaling behavior of the wave function, since the IPR is an averaged quantity over space\cite{PhysRevB.83.184206,PhysRevB.96.045138}. Thus, we study the system size dependence of the IPR, which gives the spatial distribution of the wave function. In particular, it is known that for sufficiently large system size $N$, the IPR exhibits scaling behavior as $N^{-D_2}$, where $D_2$ is called the fractal dimension\cite{Shen1994,PhysRevB.105.045146,PhysRevB.96.045138}. An exponentially localized state has $D_2=0$ and a uniformly extended state has $D_2=1$. The critical states have the intermediate fractal dimensions $0<D_2<1$\cite{PhysRevB.105.045146,PhysRevB.96.045138}.

\begin{figure*}[t]
\centering
\includegraphics[width=1\textwidth]{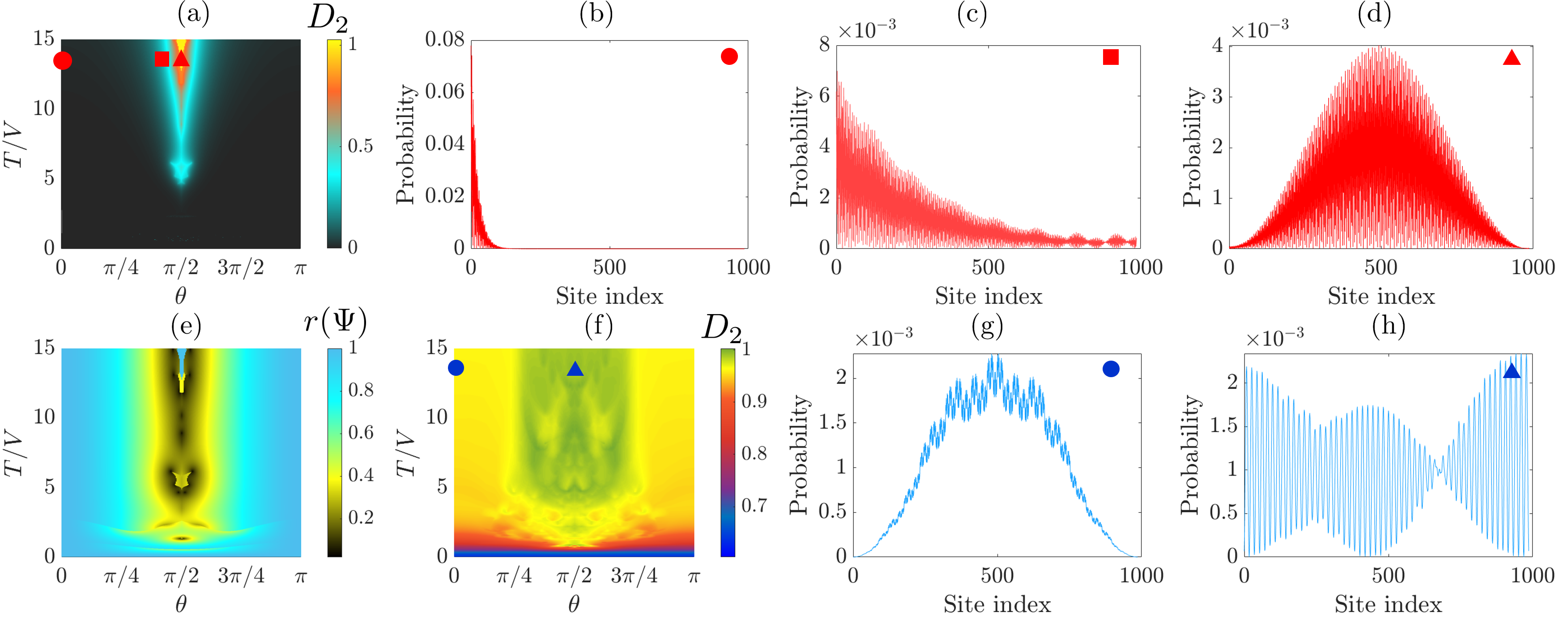}
\caption{\label{fig: result} Change of the localization characteristics of the states in the Fibonacci quasicrystal. The landscape of fractal dimensions of the states for (a) maximum value of IPR and (f) minimum value of IPR, corresponding to the maximally localized and maximally extended state, respectively. (b-d) Evolution of the localization characteristics of the maximally localized state for (b) $\theta=0$, (c) $\theta=17\pi/36$ and (d) $\theta=\pi/2$, respectively, with $T=13V$. $V=(V_A-V_B)/2$ is the difference between two kinds of the on-site energies, $V_A$ and $V_B$ in the Fibonacci quasicrystal model. Here, $V_A=1$ and $V_B=-1$. The fractal dimensions are (b) $D_2=0$, (c) $D_2=0.411$ and (d) $D_2=1$. As the phase approaches the maximum value of non-Hermiticity at $\theta=\pi/2$, the localization characteristics of the states change from the exponentially localized state (b) to the extended state (d), where intermediate power-law decaying critical states (c) appear at intermediate $\theta$. (e) The phase rigidity of the maximally localized state. Along the boundary where the fractal dimension changes rapidly in panel (a), the phase rigidity also drops steeply in panel (e). (g-h) Evolution of the localization characteristics of the maximally extended state for (g) $\theta=0$ and (h) $\theta=\pi/2$, respectively, with $T=13V$. The fractal dimensions are (g) $D_2=0.915$ and (h) $D_2=1$. The localization properties change from the self-similar critical state (g) for the Hermitian case to the uniformly oscillating extended state (h) for the maximally non-Hermitian case. See the main text for further details.}
\end{figure*}

Fig.\ref{fig: result}(a) and (f) show the landscapes of the fractal dimension of the maximally localized state and the maximally extended state, respectively, as a function of the magnitude and phase of the hopping parameter $T$ and $\theta$. The potential difference between atoms $A$ and $B$ is given by $V_A-V_B=2V$. From Eq.\eqref{AAF potential}, $V=\lambda$. As $\theta$ approaches $\pi/2$, the fractal dimensions of the maximally localized or maximally extended states increase, indicating the delocalization of the states.
In detail, Fig.\ref{fig: result} (b)-(d) show that for sufficiently large $T$ ($T>10V$ in Fig.\ref{fig: result} (a)), how the localization characteristics of the maximally localized state are controlled from the exponentially localized to the sinusoidally extended one in terms of $\theta$.
In the Hermitian case, the fractal dimension of the maximally localized state remains zero, corresponding to the exponentially localized state, regardless of the value of $T$ (see Fig.\ref{fig: result} (a) and (b)). However, in the non-Hermitian cases, the fractal dimension of the maximally localized state becomes non-zero. 
Fig.\ref{fig: result} (c) shows the wave function of the maximally localized state for $\theta=17\pi/36$ with $T=13V$ (red square in Fig.\ref{fig: result} (a)). In this case, the wave function shows the critical state with a power law decay, and the fractal dimension $D_2 =0.411$.
Moreover, Fig.\ref{fig: result} (d) shows the perfectly delocalized wave function with $D_2=1$ for $\theta=\pi/2$ with $T=13V$ (red triangle in Fig.\ref{fig: result} (a)). It is surprising that for sufficiently large but finite hopping magnitude, the maximally localized state become an extended state with $D_2=1$.
In this case, every eigenstate has $D_2=1$, which is never allowed in the Hermitian system for any finite hopping magnitude $T$ and non-zero $V$.

The reason why we can universally achieve $D_2=1$ even with finite hopping magnitude in the Fibonacci quasicrystal is as follows. The non-reciprocal hopping phase controls the localization of the states by forming a non-trivial interference between the hopping and the potential contributions of the state, which arise from the hopping parameters and the spatial potential gradient, respectively. If $T$ is sufficiently large, the interference becomes destructive as $\theta$ approaches $\pi/2$. Consequently, the potential contribution can be canceled out by the hopping contribution. As a result, the non-Hermiticity leads to the delocalization of the states by effectively blinding the spatial potential gradient such as the quasiperiodic structures with the destructive interference.
This allows us to achieve the uniformly extended state even in the presence of the spatial potential gradient with finite hopping magnitudes.

Although the maximally localized state remains exponentially localized as $D_2=0$ regardless of $\theta$ when $T$ is small ($T\lesssim 4V$ in Fig.\ref{fig: result} (a)), the non-Hermiticity increases the localization length of the state. It turns out that the interference arising from the non-reciprocal hopping phase induces the penetration of the wave function into the bulk of the system.
Thus, even with a small $T$, we can manipulate the localization strength of the state with respect to the non-reciprocal hopping phase. See Appendix \ref{sec: A2} for detailed information on controlling the localization length.

The strong state hybridization and coalescence are also important for understanding the change in the scaling properties of the state. To capture this, we compute the phase rigidity, $r(\Psi)$ of the maximally localized state. Remind that $r(\Psi)=1$ for the Hermitian case, while $r(\Psi)\le 1$ for the non-Hermitian case, because the right eigenstates could be non-orthonormal to each other. Fig.\ref{fig: result} (e) shows the landscape of phase rigidity. Comparing Fig.\ref{fig: result} (a) and (e), the phase rigidity suddenly drops at the boundaries where the change in localization characteristics occurs. Thus, when the localization characteristics change, the maximally localized state strongly hybridizes with other critical or extended states by passing through the exceptional point\cite{PhysRevX.6.021007}. This leads to delocalization of the scaling behavior of the maximally localized state.

Now let us take a look at the maximally extended state. In the Hermitian system, the maximally extended state has the fractal dimension $D_2<1$ for any finite $T$ due to the fractal structure of the Fibonacci quasicrystal\cite{jullien2012universalities}. However, Fig.\ref{fig: result} (f) shows that the non-Hermiticity can increase the fractal dimension of the maximally extended state as $D_2=1$. 
Comparing the wave functions for the maximally extended states at $\theta=0$ (Fig.\ref{fig: result} (g)) and $\pi/2$ (Fig.\ref{fig: result} (h)), one can see {\it the disappearence of the fractality} in the wave function due to the non-Hermiticity. It turns out that the non-reciprocal phase factor of the hopping parameters induces delocalization by shielding the detailed structure of the lattice, such as the Fibonacci pattern, through the interference effect, rather than simply increasing the mobility of the particle.

\begin{figure}[]
\centering
\includegraphics[width=0.6\textwidth]{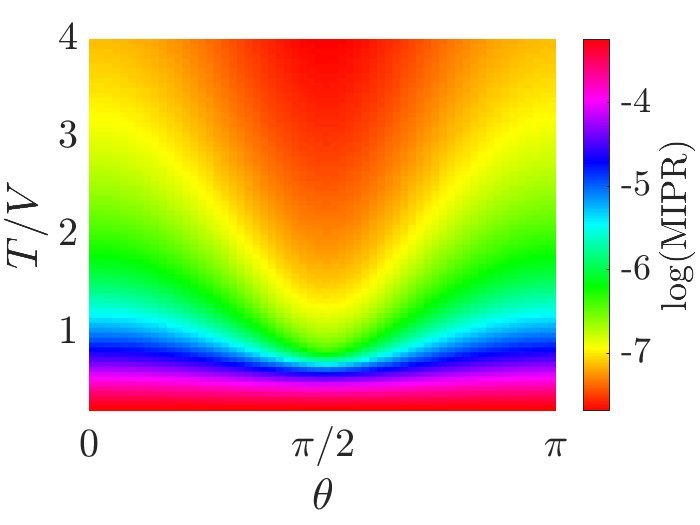}
\caption{\label{fig: MIPR} The mean value of the IPR (MIPR) of the energy spectrum as the function of the strength of the non-Hermiticity given by the phase angle of the hopping parameter, $\theta$ in the Fibonacci quasicrystal model. At $\theta=\pi/2$, the non-Hermiticity becomes maximum for given $T/V$, where $T$ is the hopping magnitude and $V=(V_A-V_B)/2$ is the difference between two kinds of the on-site energies, $V_A$ and $V_B$ in the Fibonacci quasicrystal model. Here, $V_A=1$ and $V_B=-1$. The MIPR (Eq.\eqref{MIPR}) which is the amount of the localization in the spectrum decreases as the non-Hermiticity becomes stronger.}
\end{figure}

One can ask how the localization strength of the other states changes as the strength of the non-Hermiticity increases. For a given finite $T\ge 0.2V$, we generally see that the MIPR decreases as $\theta$ gets closer to $\pi/2$ in the Fibonacci quasicrystal (See Fig.\ref{fig: MIPR}). Thus, non-Hermiticity leads to the delocalization of states.
In detail, Fig.\ref{fig: MIPR} shows the MIPR as a function of the phase of the hopping parameter, $\theta$, for different hopping magnitudes $T$ in the Fibonacci quasicrystal. For the general $T$, the MIPR decreases as $\theta$ gets closer to $\pi/2$. Thus, the localization strength in the spectrum is suppressed in the Fibonacci quasicrystal due to the non-Hermiticity. Note that in the Fibonacci quasicrystal, even for the small $T$ regime, most of the states are critical states, which are delocalized as a power-law scaling. Thus, the hybridization of the eigenstates of the Hermitian Hamiltonian due to the non-reciprocal hopping phase occurs mainly between pairs of critical states in a way to compensate the power-law decaying probability amplitudes. This gives rise to the general delocalization tendency in terms of the non-reciprocal hopping phase.

\section{Discussion and Conclusion}
\label{sec:discussion}


Over the past decade, advances in photonics and electronics have made it possible to experimentally realize open systems, described as various effective non-Hermitian Hamiltonians. 
However, how non-Hermiticity controls localization properties of the wavefunctions, is hardly understood.
In this work, we show that the localization of wavefunctions, including quasi-periodic structures and disordered chains, is tuned in non -Hermitian system with non-reciprocal hopping phase. We show that the non-Hermiticity imposed by the non-reciprocal hopping phase does not cause a skin effect, but can dramatically change wave function localization, even inducing general delocalization with sufficiently strong non-Hermiticity. While the non-Hermitian skin effect depends strongly on the boundary conditions, the delocalization induced by the non-reciprocal hopping phase does not depend on the specific boundary conditions. We show that there are two main mechanisms by which the non-reciprocal hopping phase changes the localization properties of the wave function.

The first mechanism is through interference between transition probability amplitudes. The non-reciprocal hopping phase imposes a nontrivial phase difference between the transition probability amplitudes given by different paths. This nontrivial phase causes destructive interference between the transition probability amplitudes, which is impossible in the Hermitian system, and results in eigenstates being able to spread further in space. The effect of this interference grows with increasing magnitude of the hopping parameter, as well as with non-reciprocal hopping phases, and leads to delocalization of the wave function in general.

Another mechanism is the coalescence of the eigenstates through the exceptional points. In terms of phase rigidity, we clarify that the localized (either exponentially or critically) state merges to the delocalized state at the boundary points where the localization characteristics given by the fractal dimension are drastically changed [cf. comparison between Fig.\ref{fig: result} (a) and (e)]. Thus, the states are delocalized when passing through the exceptional point. Remarkably, this is different from the case of the skin effect, where the presence of the exceptional point leads to macroscopic localization. Thus, our study extends the role of non-Hermitian exceptional points from the traditional skin effect to drastically changing the fractal dimension of the states.

We provide a novel way to control the localization of quantum states by using non-Hermiticity. Our non-Hermitian model, whose non-Hermiticity is originated from the non-reciprocal phase of the hopping parameter, has no specific directional preference. It is important to note that, depending on the state, the localization can be enhanced or reduced as a function of the non-reciprocal hopping phase and the magnitude of the hopping parameter. In this way, one can finely control over the localization of the states. In particular, for exponentially localized states or critical states present in quasicrystalline systems, the non-Hermiticity induces a perfect delocalization of the states, resulting in {\it the disappearence of the fractality}. Using the Fibonacci quasicrystal as an example, we have shown that non-Hermiticity can indeed change the localization characteristics between localized, critical and uniformly extended states. Again, this is due to the interference between transition amplitudes with respect to the non-reciprocal hopping phase and hybridization of the states through the exceptional points, In this way, they compensate for the non-uniform amplitudes of the probabilities and lead to the delocalization of the states in a regime of strong non-Hermiticity. Our work opens the utility of non-Hermiticity for high controllability of the localization characteristics.

Our theoretical work could be studied by the photonic crystal\cite{PhysRevB.104.125416} or electrical circuits similar to other open system models governed by the Lindblad master equation or the effective non-Hermitian Hamiltonian. In particular, we suggest an experimental setup to demonstrate the control of the localization of the wave function in the quasiperiodic system as shown in Fig.\ref{fig: impact}. By changing the phase accumulated by the phase shifter, one can explore the different localization characteristics of the wave functions from the exponentially localized to the critical or extended states. The electric circuit or acoustic lattice are also possible platforms to demonstrate localization control in terms of the non-reciprocal phase of the transporting waves\cite{gu2022transient}. Thus, our work helps to exploit the different localization properties of the wave functions in experiments such as quantum transport.

While we consider the one-dimensional systems, as an interesting future work, one can generalize our model to higher-dimensional systems such as two- or three-dimensional lattices\cite{PhysRevB.102.241202,PhysRevB.102.205118,PhysRevB.106.L161401}, or even to the multi-frequency driven Floquet systems with synthesized dimensions\cite{PhysRevX.13.011003,PhysRevA.107.043309}. We suspect that similar delocalization phenomena would occur in the higher-dimensional lattices due to the non-reciprocal phases of the hopping parameter. In such a case,  the state-dependent control of localization strength and the emergence of subdimensional fractality can also be discussed.

\section*{Acknowledgements}
We thank Yidong Chong and Dung Nguyen Xuan for useful discussions.


\paragraph{Funding information}
J.M.J. and S.B.L. are supported by National Research Foundation Grant (No. 2021R1A2C1093060)).

\begin{appendix}

\section{Control of localization in the uniformly random disordered chain}
\label{sec: A1}

\begin{figure}[]
\centering
\includegraphics[width=0.7\textwidth]{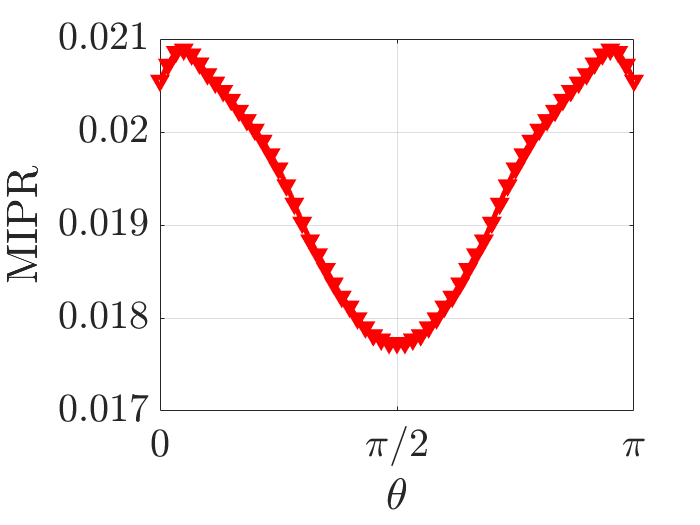}
\caption{\label{fig: supprand} MIPR as a function of the phase of the non-Hermitian hopping parameter, $\theta$, in the randomly disordered system. The degree of disorder of the on-site potential energy is $50\%$. The system size, $N=233$, and the hopping parameter value, $T=4V=4$.}
\end{figure}

Here we consider the randomly disordered chain where the on-site energies have a $50\%$ level of disorder. Specifically, the random on-site energies are between $-1.5V$ and $-0.5V$. Figure.\ref{fig: supprand} shows that the localization is suppressed as the non-Hermicity increases even for the randomly disordered system. The MIPR decreases as $\theta$ becomes $\pi/2$, which corresponds to the maximum strength of the non-Hermiticity for a given hopping magnitude $T$. Thus, the delocalization is induced by the non-reciprocal hopping phases in the randomly disordered chain system.

\section{Localization length change in the Fibonacci quasicrystal with small hopping magnitude regime}
\label{sec: A2}
\begin{figure}[]
\centering
\includegraphics[width=0.7\textwidth]{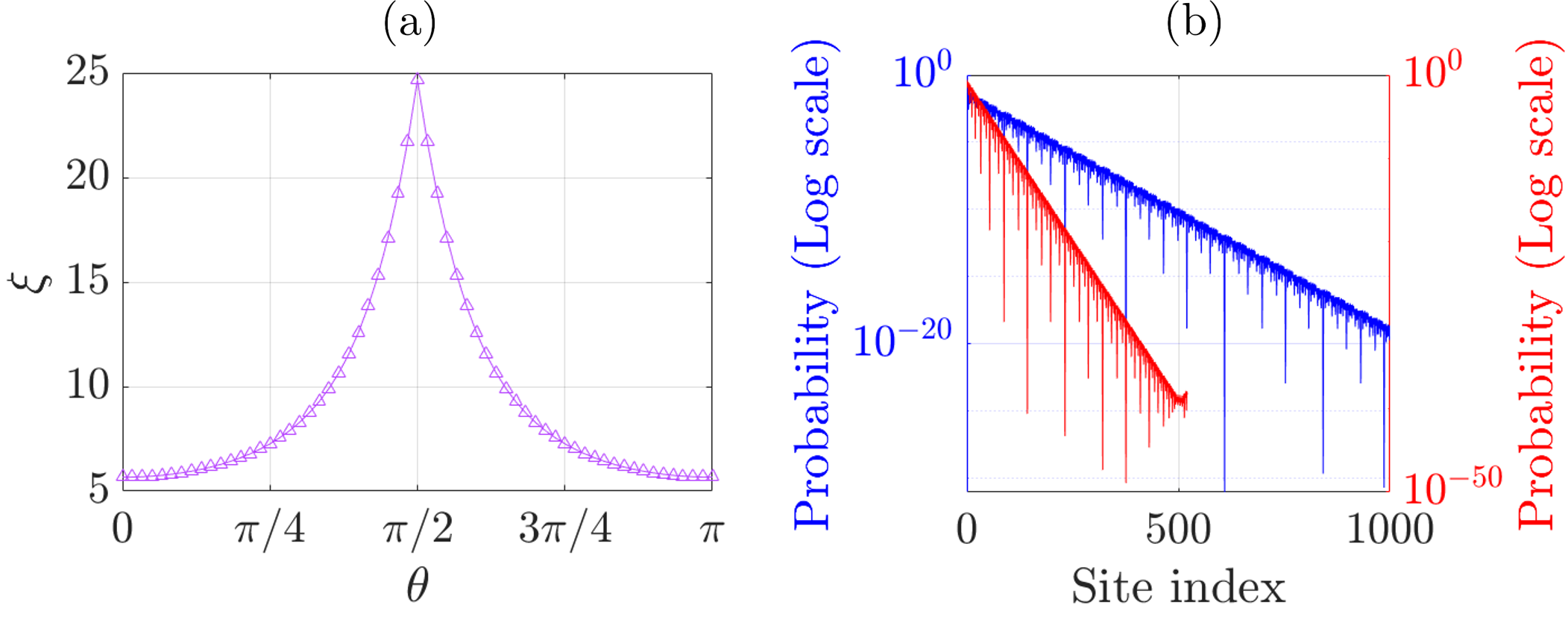}
\caption{\label{fig: supp} (a) The localization length ($\xi$ defined in Eq.\eqref{localization legnth}) changes as a function of the non-reciprocal hopping phase ($\theta$) in the Fibonacci quasicrystal model. Here, the unit of the localization length is the atomic spacing between neighboring atoms, which is set to be 1. The non-Hermiticity induces the delocalization, so the localization length increases as the non-Hermiticity becomes stronger. (b) Comparison of the probability distribution of the maximally localized states for (blue) $\theta=\pi/2$ and (red) $\theta=0$ in the logarithmic scale. The linear scaling in the figure indicates the exponential decay. The smaller slope indicates the larger localization length for the non-Hermitian case. The hopping magnitude is $T=3V=3$. The system size is $N=987$.}
\end{figure}

Here we consider the Fibonacci quasicrystal with the small hopping magnitude, $T$ regime. The localization length is controlloed by the strength of the non-Hermiticity, although the fractal dimension is always zero for the maximally localized state with the maximum value of the IPR in the small $T$ regime. In detail, let us define the localization length, $\xi$, of the exponentially localized state as follows.
\begin{align}
\label{localization legnth}
&\xi=\sqrt{\braket{\hat{x}^2}-\braket{\hat{x}}^2}
\end{align}
where $\hat{x}$ is the position operator and $\braket{\hat{O}}$ is the expectation value of the operator, $\hat{O}$. For $T=3V$ we show the variance of the localization length as a function of the non-reciprocal phase of the hopping parameter, $\theta$. Recall that the uniform non-Hermitian hopping parameter is given by $t=Te^{i\theta}$. Fig.\ref{fig: supp} (a) shows that the localization length increases drastically as $\theta$ approaches $\pi/2$. Thus, although the localization characteristics of the maximally localized state are exponentially decaying for any $\theta$ with small $T$, the localization length could be manipulated in terms of the non-reciprocal hopping phase. Fig.\ref{fig: supp} (b) compares the probability distribution of the maximally localized states for $\theta=0$ (red) and $\theta=\pi/2$ (blue) in the logarithmic scale. The different slopes show that in the non-Hermitian case the wave is able to penetrate more into the bulk.


\end{appendix}

\newpage



\bibliography{my1.bib}

\begin{thebibliography}{10}
\providecommand{\url}[1]{\texttt{#1}}
\providecommand{\urlprefix}{URL }
\expandafter\ifx\csname urlstyle\endcsname\relax
  \providecommand{\doi}[1]{doi:\discretionary{}{}{}#1}\else
  \providecommand{\doi}{doi:\discretionary{}{}{}\begingroup
  \urlstyle{rm}\Url}\fi
\providecommand{\eprint}[2][]{\url{#2}}

\bibitem{jaric2012introduction}
M.~V. Jaric,
\newblock \emph{Introduction to the Mathematics of Quasicrystals},
\newblock Elsevier,
\newblock ISBN 978-0124336315 (2012).

\bibitem{belin2000quasicrystals}
E.~Belin-Ferr,
\newblock \emph{Quasicrystals: Current Topics: Aussois, France, 16-21 May
  1999},
\newblock World Scientific,
\newblock ISBN 9810242816 (2000).

\bibitem{suck2013quasicrystals}
J.-B. Suck, M.~Schreiber and P.~H{\"a}ussler,
\newblock \emph{Quasicrystals: An introduction to structure, physical
  properties and applications}, vol.~55,
\newblock Springer Science \& Business Media,
\newblock ISBN 978-3540642244 (2013).

\bibitem{janssen2018aperiodic}
T.~Janssen, G.~Chapuis and M.~De~Boissieu,
\newblock \emph{Aperiodic crystals: from modulated phases to quasicrystals:
  structure and properties},
\newblock Oxford University Press,
\newblock ISBN 978-0198824442 (2018).

\bibitem{macia2020quasicrystals}
E.~Maci{\'a}-Barber,
\newblock \emph{Quasicrystals: fundamentals and applications},
\newblock CRC Press,
\newblock ISBN 978-0367678937 (2020).

\bibitem{jaric1986diffraction}
M.~V. Jari{\'c},
\newblock \emph{Diffraction from quasicrystals: Geometric structure factor},
\newblock Physical Review B \textbf{34}(7), 4685 (1986),
\newblock \doi{10.1103/PhysRevB.34.4685}.

\bibitem{poon1992electronic}
S.~Poon,
\newblock \emph{Electronic properties of quasicrystals an experimental review},
\newblock Advances in Physics \textbf{41}(4), 303 (1992),
\newblock \doi{10.1080/00018739200101513}.

\bibitem{goldman1991quasicrystal}
A.~Goldman and M.~Widom,
\newblock \emph{Quasicrystal structure and properties},
\newblock Annual Review of Physical Chemistry \textbf{42}(1), 685 (1991),
\newblock \doi{10.1146/annurev.pc.42.100191.003345}.

\bibitem{sarkar2021anderson}
S.~Sarkar, M.~Kraj{\v{c}}{\'\i}, P.~Sadhukhan, V.~K. Singh, A.~Gloskovskii,
  P.~Mandal, V.~Fourn{\'e}e, M.-C. de~Weerd, J.~Ledieu, I.~R. Fisher
  \emph{et~al.},
\newblock \emph{Anderson localization of electron states in a quasicrystal},
\newblock Physical Review B \textbf{103}(24), L241106 (2021),
\newblock \doi{10.1103/PhysRevB.103.L241106}.

\bibitem{rosa2021exploring}
M.~I. Rosa, Y.~Guo and M.~Ruzzene,
\newblock \emph{Exploring topology of 1{D} quasiperiodic metastructures through
  modulated {LEGO} resonators},
\newblock Applied Physics Letters \textbf{118}(13), 131901 (2021),
\newblock \doi{10.1063/5.0042294}.

\bibitem{xiao2021observation}
T.~Xiao, D.~Xie, Z.~Dong, T.~Chen, W.~Yi and B.~Yan,
\newblock \emph{Observation of topological phase with critical localization in
  a quasi-periodic lattice},
\newblock Science Bulletin \textbf{66}(21), 2175 (2021),
\newblock \doi{10.1016/j.scib.2021.07.025}.

\bibitem{PhysRevResearch.3.013168}
J.~Jeon and S.~Lee,
\newblock \emph{Topological critical states and anomalous electronic
  transmittance in one-dimensional quasicrystals},
\newblock Phys. Rev. Res. \textbf{3}, 013168 (2021),
\newblock \doi{10.1103/PhysRevResearch.3.013168}.

\bibitem{deguchi2012quantum}
K.~Deguchi, S.~Matsukawa, N.~K. Sato, T.~Hattori, K.~Ishida, H.~Takakura and
  T.~Ishimasa,
\newblock \emph{Quantum critical state in a magnetic quasicrystal},
\newblock Nature materials \textbf{11}(12), 1013 (2012),
\newblock \doi{10.1038/nmat3432}.

\bibitem{PhysRevB.35.1020}
M.~Kohmoto, B.~Sutherland and C.~Tang,
\newblock \emph{Critical wave functions and a {C}antor-set spectrum of a
  one-dimensional quasicrystal model},
\newblock Phys. Rev. B \textbf{35}, 1020 (1987),
\newblock \doi{10.1103/PhysRevB.35.1020}.

\bibitem{Tuz:09}
V.~R. Tuz,
\newblock \emph{Optical properties of a quasi-periodic generalized {F}ibonacci
  structure of chiral and material layers},
\newblock J. Opt. Soc. Am. B \textbf{26}(4), 627 (2009),
\newblock \doi{10.1364/JOSAB.26.000627}.

\bibitem{PhysRevB.96.045138}
N.~Mac\'e, A.~Jagannathan, P.~Kalugin, R.~Mosseri and F.~Pi\'echon,
\newblock \emph{Critical eigenstates and their properties in one- and
  two-dimensional quasicrystals},
\newblock Phys. Rev. B \textbf{96}, 045138 (2017),
\newblock \doi{10.1103/PhysRevB.96.045138}.

\bibitem{PhysRevLett.58.2436}
M.~Kohmoto, B.~Sutherland and K.~Iguchi,
\newblock \emph{Localization of optics: Quasiperiodic media},
\newblock Phys. Rev. Lett. \textbf{58}, 2436 (1987),
\newblock \doi{10.1103/PhysRevLett.58.2436}.

\bibitem{PhysRevB.101.174203}
T.~Cookmeyer, J.~Motruk and J.~E. Moore,
\newblock \emph{Critical properties of the ground-state
  localization-delocalization transition in the many-particle {A}ubry-{A}ndr\'e
  model},
\newblock Phys. Rev. B \textbf{101}, 174203 (2020),
\newblock \doi{10.1103/PhysRevB.101.174203}.

\bibitem{PhysRevB.107.054206}
A.~Jagannathan and M.~Tarzia,
\newblock \emph{Electronic states of a disordered two-dimensional quasiperiodic
  tiling: From critical states to {A}nderson localization},
\newblock Phys. Rev. B \textbf{107}, 054206 (2023),
\newblock \doi{10.1103/PhysRevB.107.054206}.

\bibitem{aubry1980analyticity}
S.~Aubry and G.~André,
\newblock \emph{Analyticity breaking and {A}nderson localization in
  incommensurate lattices},
\newblock Proceedings, VIII International Colloquium on Group-Theoretical
  Methods in Physics \textbf{3} (1980),
\newblock \doi{10.1016/j.anihpb.2003.04.002}.

\bibitem{Shen1994}
B.~Shen,
\newblock \emph{Fractal Structure of Quasicrystals}, pp. 337--348,
\newblock Springer Berlin Heidelberg, Berlin, Heidelberg,
\newblock ISBN 978-3-662-07304-9,
\newblock \doi{10.1007/978-3-662-07304-9_25} (1994).

\bibitem{PhysRevLett.115.180401}
V.~Mastropietro,
\newblock \emph{Localization of interacting {F}ermions in the {A}ubry-{A}ndr\'e
  model},
\newblock Phys. Rev. Lett. \textbf{115}, 180401 (2015),
\newblock \doi{10.1103/PhysRevLett.115.180401}.

\bibitem{PhysRevLett.97.026601}
G.~Trambly~de Laissardi\`ere, J.-P. Julien and D.~Mayou,
\newblock \emph{Quantum transport of slow charge carriers in quasicrystals and
  correlated systems},
\newblock Phys. Rev. Lett. \textbf{97}, 026601 (2006),
\newblock \doi{10.1103/PhysRevLett.97.026601}.

\bibitem{PhysRevX.6.011016}
M.~A. Bandres, M.~C. Rechtsman and M.~Segev,
\newblock \emph{Topological photonic quasicrystals: Fractal topological
  spectrum and protected transport},
\newblock Phys. Rev. X \textbf{6}, 011016 (2016),
\newblock \doi{10.1103/PhysRevX.6.011016}.

\bibitem{PhysRevB.106.134431}
J.~Jeon, S.~K. Kim and S.~Lee,
\newblock \emph{Fractalized magnon transport on a quasicrystal with enhanced
  stability},
\newblock Phys. Rev. B \textbf{106}, 134431 (2022),
\newblock \doi{10.1103/PhysRevB.106.134431}.

\bibitem{PhysRevB.102.075107}
M.~Coppolaro, G.~Castaldi and V.~Galdi,
\newblock \emph{Anomalous light transport induced by deeply subwavelength
  quasiperiodicity in multilayered dielectric metamaterials},
\newblock Phys. Rev. B \textbf{102}, 075107 (2020),
\newblock \doi{10.1103/PhysRevB.102.075107}.

\bibitem{PhysRevLett.126.145501}
E.~Cherkaev, F.~Guevara~Vasquez, C.~Mauck, M.~Prisbrey and B.~Raeymaekers,
\newblock \emph{Wave-driven assembly of quasiperiodic patterns of particles},
\newblock Phys. Rev. Lett. \textbf{126}, 145501 (2021),
\newblock \doi{10.1103/PhysRevLett.126.145501}.

\bibitem{PhysRevB.88.201404}
S.~S. Kruk, C.~Helgert, M.~Decker, I.~Staude, C.~Menzel, C.~Etrich,
  C.~Rockstuhl, C.~Jagadish, T.~Pertsch, D.~N. Neshev and Y.~S. Kivshar,
\newblock \emph{Optical metamaterials with quasicrystalline symmetry:
  Symmetry-induced optical isotropy},
\newblock Phys. Rev. B \textbf{88}, 201404 (2013),
\newblock \doi{10.1103/PhysRevB.88.201404}.

\bibitem{PhysRevResearch.4.043030}
D.~Beli, M.~I.~N. Rosa, C.~De~Marqui and M.~Ruzzene,
\newblock \emph{Wave beaming and diffraction in quasicrystalline elastic
  metamaterial plates},
\newblock Phys. Rev. Res. \textbf{4}, 043030 (2022),
\newblock \doi{10.1103/PhysRevResearch.4.043030}.

\bibitem{PhysRevLett.125.200604}
M.~Sbroscia, K.~Viebahn, E.~Carter, J.-C. Yu, A.~Gaunt and U.~Schneider,
\newblock \emph{Observing localization in a 2{D} quasicrystalline optical
  lattice},
\newblock Phys. Rev. Lett. \textbf{125}, 200604 (2020),
\newblock \doi{10.1103/PhysRevLett.125.200604}.

\bibitem{stadnik1998physical}
Z.~M. Stadnik,
\newblock \emph{Physical properties of quasicrystals}, vol. 126,
\newblock Springer Science \& Business Media,
\newblock \doi{10.1007/978-3-642-58434-3} (1998).

\bibitem{PhysRevX.13.021007}
K.~Kawabata, T.~Numasawa and S.~Ryu,
\newblock \emph{Entanglement {P}hase {T}ransition induced by the
  {N}on-{H}ermitian {S}kin {E}ffect},
\newblock Phys. Rev. X \textbf{13}, 021007 (2023),
\newblock \doi{10.1103/PhysRevX.13.021007}.

\bibitem{10.1063/1.5115323}
D.~Manzano,
\newblock \emph{{A short introduction to the Lindblad master equation}},
\newblock AIP Advances \textbf{10}(2), 025106 (2020),
\newblock \doi{10.1063/1.5115323}.

\bibitem{niu2023effect}
X.~Niu, J.~Li, S.~L. Wu and X.~X. Yi,
\newblock \emph{Effect of quantum jumps on non-hermitian system} (2023),
  \eprint{2202.12591}.

\bibitem{gao2015observation}
T.~Gao, E.~Estrecho, K.~Bliokh, T.~Liew, M.~Fraser, S.~Brodbeck, M.~Kamp,
  C.~Schneider, S.~H{\"o}fling, Y.~Yamamoto \emph{et~al.},
\newblock \emph{Observation of non-{H}ermitian degeneracies in a chaotic
  exciton-polariton billiard},
\newblock Nature \textbf{526}(7574), 554 (2015),
\newblock \doi{10.1038/nature15522}.

\bibitem{PhysRevLett.125.123902}
S.~Mandal, R.~Banerjee, E.~A. Ostrovskaya and T.~C.~H. Liew,
\newblock \emph{Nonreciprocal transport of exciton polaritons in a
  non-hermitian chain},
\newblock Phys. Rev. Lett. \textbf{125}, 123902 (2020),
\newblock \doi{10.1103/PhysRevLett.125.123902}.

\bibitem{PhysRevB.104.235408}
Z.-F. Yu, J.-K. Xue, L.~Zhuang, J.~Zhao and W.-M. Liu,
\newblock \emph{Non-hermitian spectrum and multistability in exciton-polariton
  condensates},
\newblock Phys. Rev. B \textbf{104}, 235408 (2021),
\newblock \doi{10.1103/PhysRevB.104.235408}.

\bibitem{wang2021topological}
H.~Wang, X.~Zhang, J.~Hua, D.~Lei, M.~Lu and Y.~Chen,
\newblock \emph{Topological physics of non-{H}ermitian optics and photonics: a
  review},
\newblock Journal of Optics \textbf{23}(12), 123001 (2021),
\newblock \doi{10.1088/2040-8986/ac2e15}.

\bibitem{carmichael2009open}
H.~Carmichael,
\newblock \emph{An open systems approach to quantum optics: lectures presented
  at the Universit{\'e} Libre de Bruxelles, October 28 to November 4, 1991},
  vol.~18,
\newblock Springer Science \& Business Media,
\newblock \doi{10.1007/978-3-540-47620-7} (2009).

\bibitem{PhysRevB.105.L180406}
K.~Deng and B.~Flebus,
\newblock \emph{Non-hermitian skin effect in magnetic systems},
\newblock Phys. Rev. B \textbf{105}, L180406 (2022),
\newblock \doi{10.1103/PhysRevB.105.L180406}.

\bibitem{hurst2022non}
H.~M. Hurst and B.~Flebus,
\newblock \emph{Non-{H}ermitian physics in magnetic systems},
\newblock Journal of Applied Physics \textbf{132}(22), 220902 (2022),
\newblock \doi{10.1063/5.0124841}.

\bibitem{zhang2022universal}
K.~Zhang, Z.~Yang and C.~Fang,
\newblock \emph{Universal non-hermitian skin effect in two and higher
  dimensions},
\newblock Nature communications \textbf{13}(1), 2496 (2022),
\newblock \doi{10.1038/s41467-022-30161-6}.

\bibitem{PhysRevB.104.125416}
J.~Zhong, K.~Wang, Y.~Park, V.~Asadchy, C.~C. Wojcik, A.~Dutt and S.~Fan,
\newblock \emph{Nontrivial point-gap topology and non-hermitian skin effect in
  photonic crystals},
\newblock Phys. Rev. B \textbf{104}, 125416 (2021),
\newblock \doi{10.1103/PhysRevB.104.125416}.

\bibitem{PhysRevB.100.054301}
H.~Jiang, L.-J. Lang, C.~Yang, S.-L. Zhu and S.~Chen,
\newblock \emph{Interplay of non-hermitian skin effects and {A}nderson
  localization in nonreciprocal quasiperiodic lattices},
\newblock Phys. Rev. B \textbf{100}, 054301 (2019),
\newblock \doi{10.1103/PhysRevB.100.054301}.

\bibitem{PhysRevB.103.054203}
S.~Longhi,
\newblock \emph{Phase transitions in a non-hermitian {A}ubry-{A}ndr\'e-{H}arper
  model},
\newblock Phys. Rev. B \textbf{103}, 054203 (2021),
\newblock \doi{10.1103/PhysRevB.103.054203}.

\bibitem{PhysRevResearch.3.033184}
L.~Zhou,
\newblock \emph{Floquet engineering of topological localization transitions and
  mobility edges in one-dimensional non-hermitian quasicrystals},
\newblock Phys. Rev. Research \textbf{3}, 033184 (2021),
\newblock \doi{10.1103/PhysRevResearch.3.033184}.

\bibitem{PhysRevLett.77.570}
N.~Hatano and D.~R. Nelson,
\newblock \emph{Localization transitions in non-hermitian quantum mechanics},
\newblock Phys. Rev. Lett. \textbf{77}, 570 (1996),
\newblock \doi{10.1103/PhysRevLett.77.570}.

\bibitem{PhysRevLett.80.5172}
N.~M. Shnerb and D.~R. Nelson,
\newblock \emph{Winding numbers, complex currents, and non-hermitian
  localization},
\newblock Phys. Rev. Lett. \textbf{80}, 5172 (1998),
\newblock \doi{10.1103/PhysRevLett.80.5172}.

\bibitem{PhysRevLett.124.086801}
N.~Okuma, K.~Kawabata, K.~Shiozaki and M.~Sato,
\newblock \emph{Topological origin of non-hermitian skin effects},
\newblock Phys. Rev. Lett. \textbf{124}, 086801 (2020),
\newblock \doi{10.1103/PhysRevLett.124.086801}.

\bibitem{zhang2021observation}
X.~Zhang, Y.~Tian, J.-H. Jiang, M.-H. Lu and Y.-F. Chen,
\newblock \emph{Observation of higher-order non-hermitian skin effect},
\newblock Nature communications \textbf{12}(1), 1 (2021),
\newblock \doi{10.1038/s41467-021-25716-y}.

\bibitem{zhou2022engineering}
L.~Zhou, H.~Li, W.~Yi and X.~Cui,
\newblock \emph{Engineering non-hermitian skin effect with band topology in
  ultracold gases},
\newblock Communications Physics \textbf{5}(1), 1 (2022),
\newblock \doi{10.1038/s42005-022-01021-y}.

\bibitem{PhysRevLett.127.256402}
M.~Lu, X.-X. Zhang and M.~Franz,
\newblock \emph{Magnetic suppression of non-hermitian skin effects},
\newblock Phys. Rev. Lett. \textbf{127}, 256402 (2021),
\newblock \doi{10.1103/PhysRevLett.127.256402}.

\bibitem{yao2018edge}
S.~Yao and Z.~Wang,
\newblock \emph{Edge states and topological invariants of non-{H}ermitian
  systems},
\newblock Physical review letters \textbf{121}(8), 086803 (2018),
\newblock \doi{10.1103/PhysRevLett.121.086803}.

\bibitem{PhysRevB.106.134112}
H.~Gao, H.~Xue, Z.~Gu, L.~Li, W.~Zhu, Z.~Su, J.~Zhu, B.~Zhang and Y.~D. Chong,
\newblock \emph{Anomalous floquet non-hermitian skin effect in a ring resonator
  lattice},
\newblock Phys. Rev. B \textbf{106}, 134112 (2022),
\newblock \doi{10.1103/PhysRevB.106.134112}.

\bibitem{doi:10.1126/science.abf6568}
K.~Wang, A.~Dutt, K.~Y. Yang, C.~C. Wojcik, J.~Vučković and S.~Fan,
\newblock \emph{Generating arbitrary topological windings of a non-{H}ermitian
  band},
\newblock Science \textbf{371}(6535), 1240 (2021),
\newblock \doi{10.1126/science.abf6568}.

\bibitem{doi:10.1126/science.aaz3071}
A.~Dutt, Q.~Lin, L.~Yuan, M.~Minkov, M.~Xiao and S.~Fan,
\newblock \emph{A single photonic cavity with two independent physical
  synthetic dimensions},
\newblock Science \textbf{367}(6473), 59 (2020),
\newblock \doi{10.1126/science.aaz3071}.

\bibitem{lee2018topolectrical}
C.~H. Lee, S.~Imhof, C.~Berger, F.~Bayer, J.~Brehm, L.~W. Molenkamp,
  T.~Kiessling and R.~Thomale,
\newblock \emph{Topolectrical circuits},
\newblock Communications Physics \textbf{1}(1), 39 (2018),
\newblock \doi{10.1038/s42005-018-0035-2}.

\bibitem{Wu:16}
B.~Wu, B.~Wu, J.~Xu, J.~Xiao and Y.~Chen,
\newblock \emph{Coupled mode theory in non-hermitian optical cavities},
\newblock Opt. Express \textbf{24}(15), 16566 (2016),
\newblock \doi{10.1364/OE.24.016566}.

\bibitem{PhysRevA.102.023501}
B.-Y. Sun and Z.-W. Zhou,
\newblock \emph{One-dimensional one-band topologically nontrivial non-hermitian
  system simulated in optical cavities},
\newblock Phys. Rev. A \textbf{102}, 023501 (2020),
\newblock \doi{10.1103/PhysRevA.102.023501}.

\bibitem{gu2022transient}
Z.~Gu, H.~Gao, H.~Xue, J.~Li, Z.~Su and J.~Zhu,
\newblock \emph{Transient non-hermitian skin effect},
\newblock Nature Communications \textbf{13}(1), 7668 (2022),
\newblock \doi{10.1038/s41467-022-35448-2}.

\bibitem{PhysRevB.100.125157}
S.~Longhi,
\newblock \emph{Metal-insulator phase transition in a non-{H}ermitian
  {A}ubry-{A}ndr\'e-{H}arper model},
\newblock Phys. Rev. B \textbf{100}, 125157 (2019),
\newblock \doi{10.1103/PhysRevB.100.125157}.

\bibitem{PhysRevB.104.224204}
Z.~Xu, X.~Xia and S.~Chen,
\newblock \emph{Non-{H}ermitian {A}ubry-{A}ndr\'e model with power-law
  hopping},
\newblock Phys. Rev. B \textbf{104}, 224204 (2021),
\newblock \doi{10.1103/PhysRevB.104.224204}.

\bibitem{PhysRevB.106.014204}
L.-J. Zhai, G.-Y. Huang and S.~Yin,
\newblock \emph{Nonequilibrium dynamics of the localization-delocalization
  transition in the non-{H}ermitian {A}ubry-{A}ndr\'e model},
\newblock Phys. Rev. B \textbf{106}, 014204 (2022),
\newblock \doi{10.1103/PhysRevB.106.014204}.

\bibitem{PhysRevResearch.2.033052}
Q.-B. Zeng and Y.~Xu,
\newblock \emph{Winding numbers and generalized mobility edges in non-hermitian
  systems},
\newblock Phys. Rev. Res. \textbf{2}, 033052 (2020),
\newblock \doi{10.1103/PhysRevResearch.2.033052}.

\bibitem{PhysRevLett.122.237601}
S.~Longhi,
\newblock \emph{Topological phase transition in non-hermitian quasicrystals},
\newblock Phys. Rev. Lett. \textbf{122}, 237601 (2019),
\newblock \doi{10.1103/PhysRevLett.122.237601}.

\bibitem{PhysRevB.103.104203}
T.~Liu, S.~Cheng, H.~Guo and G.~Xianlong,
\newblock \emph{Fate of {M}ajorana zero modes, exact location of critical
  states, and unconventional real-complex transition in non-hermitian
  quasiperiodic lattices},
\newblock Phys. Rev. B \textbf{103}, 104203 (2021),
\newblock \doi{10.1103/PhysRevB.103.104203}.

\bibitem{PhysRevResearch.3.033257}
A.~\ifmmode~\check{S}\else \v{S}\fi{}trkalj, E.~V.~H. Doggen, I.~V. Gornyi and
  O.~Zilberberg,
\newblock \emph{Many-body localization in the interpolating
  {A}ubry-{A}ndr\'e-{F}ibonacci model},
\newblock Phys. Rev. Res. \textbf{3}, 033257 (2021),
\newblock \doi{10.1103/PhysRevResearch.3.033257}.

\bibitem{trefil2001encyclopedia}
J.~Trefil,
\newblock \emph{The encyclopedia of science and technology},
\newblock Routledge,
\newblock ISBN 978-0-12-227410-7 (2001).

\bibitem{PhysRevX.6.021007}
K.~Ding, G.~Ma, M.~Xiao, Z.~Q. Zhang and C.~T. Chan,
\newblock \emph{Emergence, coalescence, and topological properties of multiple
  exceptional points and their experimental realization},
\newblock Phys. Rev. X \textbf{6}, 021007 (2016),
\newblock \doi{10.1103/PhysRevX.6.021007}.

\bibitem{PhysRevA.95.022117}
H.~Eleuch and I.~Rotter,
\newblock \emph{Resonances in open quantum systems},
\newblock Phys. Rev. A \textbf{95}, 022117 (2017),
\newblock \doi{10.1103/PhysRevA.95.022117}.

\bibitem{PhysRevB.83.184206}
N.~C. Murphy, R.~Wortis and W.~A. Atkinson,
\newblock \emph{Generalized inverse participation ratio as a possible measure
  of localization for interacting systems},
\newblock Phys. Rev. B \textbf{83}, 184206 (2011),
\newblock \doi{10.1103/PhysRevB.83.184206}.

\bibitem{calixto2015inverse}
M.~Calixto and E.~Romera,
\newblock \emph{Inverse participation ratio and localization in topological
  insulator phase transitions},
\newblock Journal of Statistical Mechanics: Theory and Experiment
  \textbf{2015}(6), P06029 (2015),
\newblock \doi{10.1088/1742-5468/2015/06/P06029}.

\bibitem{https://doi.org/10.48550/arxiv.2112.14783}
J.~Jeon and S.~Lee,
\newblock \emph{Discovery of new quasicrystals from translation of hypercubic
  lattice},
\newblock \doi{10.48550/ARXIV.2112.14783} (2021).

\bibitem{PhysRevB.105.064502}
J.~Jeon and S.~Lee,
\newblock \emph{Pattern-dependent proximity effect and {M}ajorana edge mode in
  one-dimensional quasicrystals},
\newblock Phys. Rev. B \textbf{105}, 064502 (2022),
\newblock \doi{10.1103/PhysRevB.105.064502}.

\bibitem{https://doi.org/10.48550/arxiv.2205.15343}
J.~Jeon and S.~Lee,
\newblock \emph{Quantum bridge states and sub-dimensional transports in
  quasicrystals},
\newblock \doi{10.48550/ARXIV.2205.15343} (2022).

\bibitem{PhysRevB.104.014202}
L.-J. Zhai, G.-Y. Huang and S.~Yin,
\newblock \emph{Cascade of the delocalization transition in a non-hermitian
  interpolating {A}ubry-{A}ndr\'e-{F}ibonacci chain},
\newblock Phys. Rev. B \textbf{104}, 014202 (2021),
\newblock \doi{10.1103/PhysRevB.104.014202}.

\bibitem{doi:10.1126/science.aar7709}
M.-A. Miri and A.~Alù,
\newblock \emph{Exceptional points in optics and photonics},
\newblock Science \textbf{363}(6422), eaar7709 (2019),
\newblock \doi{10.1126/science.aar7709}.

\bibitem{fujiwara2007quasicrystals}
T.~Fujiwara and Y.~Ishii,
\newblock \emph{Quasicrystals},
\newblock Elsevier,
\newblock ISBN 9780080555973 (2007).

\bibitem{jullien2012universalities}
R.~Jullien, L.~Peliti, R.~Rammal and N.~Boccara,
\newblock \emph{Universalities in Condensed Matter: Proceedings of the
  Workshop, Les Houches, France, March 15--25, 1988}, vol.~32,
\newblock Springer Science \& Business Media,
\newblock \doi{10.1007/978-3-642-51005-2} (2012).

\bibitem{forrest2002topological}
A.~Forrest, J.~Hunton and J.~Kellendonk,
\newblock \emph{Topological invariants for projection method patterns}, vol.
  758,
\newblock American Mathematical Soc.,
\newblock ISBN 978-1-4704-0351-5 (2002).

\bibitem{PhysRevB.105.045146}
J.~Jeon, M.~J. Park and S.~Lee,
\newblock \emph{Length scale formation in the {L}andau levels of
  quasicrystals},
\newblock Phys. Rev. B \textbf{105}, 045146 (2022),
\newblock \doi{10.1103/PhysRevB.105.045146}.

\bibitem{PhysRevB.102.241202}
R.~Okugawa, R.~Takahashi and K.~Yokomizo,
\newblock \emph{Second-order topological non-hermitian skin effects},
\newblock Phys. Rev. B \textbf{102}, 241202 (2020),
\newblock \doi{10.1103/PhysRevB.102.241202}.

\bibitem{PhysRevB.102.205118}
K.~Kawabata, M.~Sato and K.~Shiozaki,
\newblock \emph{Higher-order non-hermitian skin effect},
\newblock Phys. Rev. B \textbf{102}, 205118 (2020),
\newblock \doi{10.1103/PhysRevB.102.205118}.

\bibitem{PhysRevB.106.L161401}
C.~C. Wojcik, K.~Wang, A.~Dutt, J.~Zhong and S.~Fan,
\newblock \emph{Eigenvalue topology of non-hermitian band structures in two and
  three dimensions},
\newblock Phys. Rev. B \textbf{106}, L161401 (2022),
\newblock \doi{10.1103/PhysRevB.106.L161401}.

\bibitem{PhysRevX.13.011003}
K.~W. Kim, D.~Bagrets, T.~Micklitz and A.~Altland,
\newblock \emph{Floquet simulators for topological surface states in
  isolation},
\newblock Phys. Rev. X \textbf{13}, 011003 (2023),
\newblock \doi{10.1103/PhysRevX.13.011003}.

\bibitem{PhysRevA.107.043309}
Y.~Wang, A.-S. Walter, G.~Jotzu and K.~Viebahn,
\newblock \emph{Topological {F}loquet engineering using two frequencies in two
  dimensions},
\newblock Phys. Rev. A \textbf{107}, 043309 (2023),
\newblock \doi{10.1103/PhysRevA.107.043309}.

\end{thebibliography}

\nolinenumbers

\end{document}